\pdfoutput=1
\documentclass[5p,10pt,number,sort&compress]{elsarticle}
\usepackage[english]{babel}
\usepackage{graphicx}
\usepackage[utf8]{inputenc}
\usepackage[T1]{fontenc}
\usepackage{fixltx2e}
\usepackage{dblfloatfix}
\usepackage{lpic}
\usepackage{url}
\usepackage{array}
\usepackage{longtable}
\usepackage{amsmath,amssymb}
\usepackage[colorlinks,citecolor=blue,filecolor=blue,linkcolor=blue,urlcolor=blue]{hyperref}
\DeclareMathOperator{\erf}{erf}
\DeclareMathOperator{\e}{e}
\newcommand{\figdir}{./}

\begin{document}

\title{Features of a fully renewable US electricity system:\\ 
  Optimized mixes of wind and solar PV and transmission grid extensions} 

\author[fias,stfd]{Sarah Becker\corref{cor}}
\ead{becker@fias.uni-frankfurt.de}

\author[stfd]{Bethany~A. Frew}

\author[aue,stfd]{Gorm~B. Andresen}

\author[aup]{Timo Zeyer}

\author[fias]{Stefan Schramm}

\author[aue,aum]{Martin Greiner}

\author[stfd]{Mark~Z. Jacobson}

\cortext[cor]{Corresponding author}

\address[fias]{
  Frankfurt Institute for Advanced Studies,
  Goethe-Universit\"at, 
  60438~Frankfurt am Main, Germany
}

\address[stfd]{
  Department of Civil and Environmental Engineering, 
  Stanford University,
  Stanford, CA, USA
}

\address[aup]{
  Department of Physics,
  Aarhus University, 
  8000~Aarhus~C, Denmark
}

\address[aue]{
  Department of Engineering,
  Aarhus University, 
  8200~Aarhus~N, Denmark
}

\address[aum]{
  Department of Mathematics,
  Aarhus University, 
  8000~Aarhus~C, Denmark
}

\begin{abstract}
  Wind and solar PV generation data for the entire contiguous US are calculated,
  on the basis of 32 years of weather data with temporal resolution of one hour and
  spatial resolution of 40$\times$40\,km$^2$, assuming site-suitability-based as
  well as stochastic wind and solar PV capacity distributions throughout the
  country. These data are used to investigate a fully renewable electricity
  system, resting primarily upon wind and solar PV power.
  We find that the seasonal optimal mix of wind and solar PV comes at around
  80\% solar PV share, owing to the US summer load peak. By picking this mix,
  long-term storage requirements can be more than halved compared to a wind only
  mix. The daily optimal mix lies at about 80\% wind share due to the nightly
  gap in solar PV production. Picking this mix instead of solar only reduces
  backup energy needs by about 50\%. Furthermore, we calculate shifts in FERC
  (Federal Energy Regulatory Commission)-level LCOE (Levelized Costs Of
  Electricity) for wind and solar PV due to their differing resource quality and
  fluctuation patterns.  LCOE vary by up to 35\% due to regional conditions, and
  LCOE-optimal mixes turn out to largely follow resource quality.
  A transmission network enhancement among FERC regions is constructed to
  transfer high penetrations of solar and wind across FERC boundaries, based on
  a novel least-cost optimization approach.
\end{abstract}


\maketitle

\section{Introduction}
\label{sec:1}

CO$_2$ and air pollution emission reduction goals as well as energy security,
price stability, and affordability considerations make renewable electricity
generation attractive. A highly renewable electricity supply will be based to a
large extent on wind and solar photovoltaic (PV) power, since these two
resources are both abundant and either relatively inexpensive or rapidly
becoming cost competitive \cite{Jacobson:2009nx}. Such a system demands a
fundamentally different design approach: While electricity generation was
traditionally constructed to be dispatchable in order to follow the demand, wind
and solar PV power output is largely determined by weather conditions that are
out of human control. We therefore collectively term them VRES (variable
renewable energy sources). Although spatial aggregation has a favorable impact
on generation characteristics as was found both for wind and solar PV power in
numerous studies \cite{archer, holttinen05, sinden07, wiemken01, mills10,
Widen:2011ys, Kempton:2010oq}, there is still a considerable mismatch between
load and generation left. 

\begin{figure}[!bt]
  \centering
  \includegraphics[width=0.49\textwidth]{\figdir/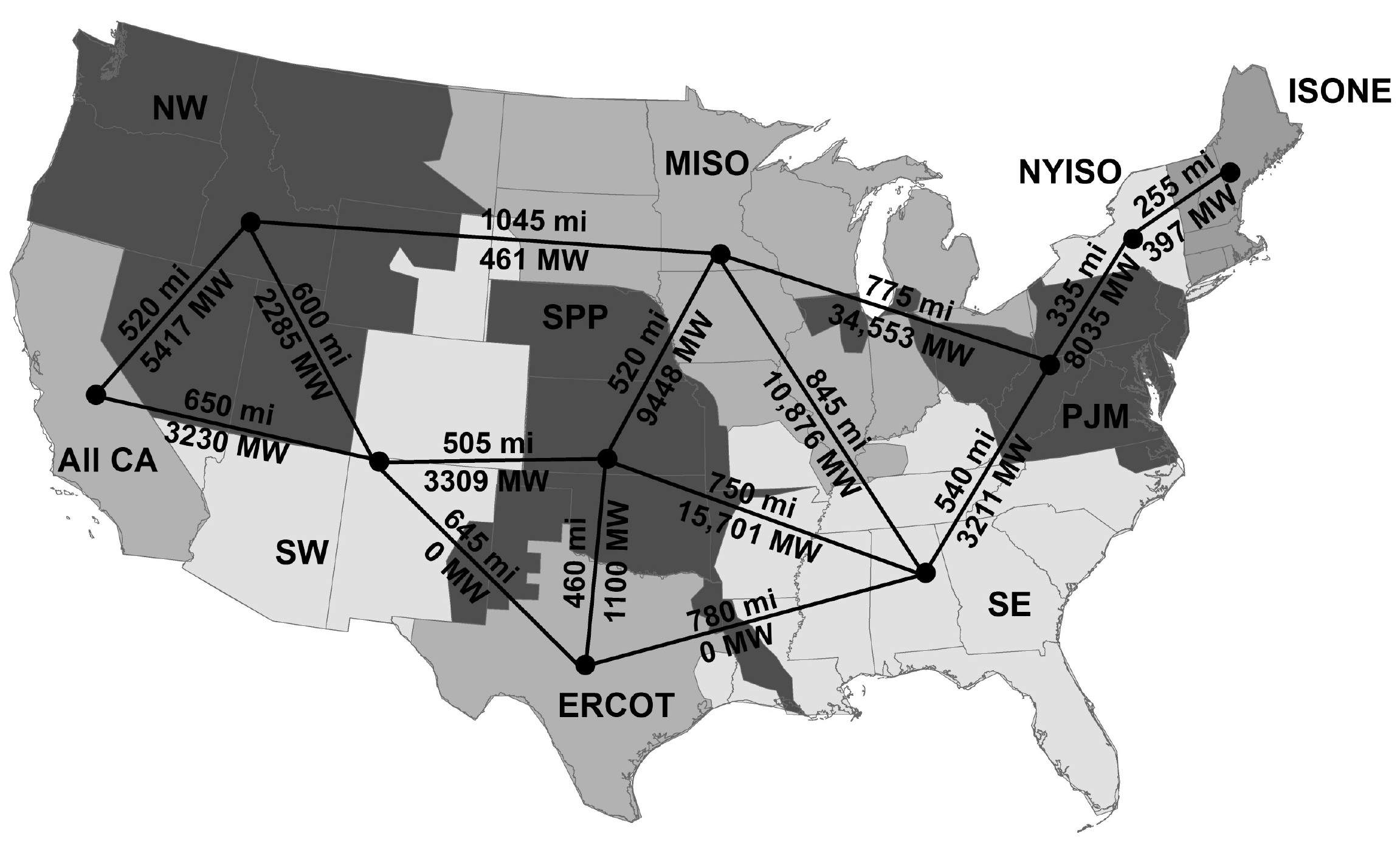}
  \caption{FERC (Federal Energy Regulatory Commission) regions in the contiguous
    US. Geographical center points, their distances, and the installed
    transmission capacity as of 2008 are included. Data as compiled in
    \cite{bethany}.
  }
  \label{fig:fercs}
\end{figure}
This paper aims to identify general design features for the US power system with
a high share of wind and solar PV. While several studies have demonstrated the
feasibility of high penetrations of VRES generators in the regional or
nationwide US electric system \cite{hart11, Budischak13, Nelson12,
NREL_re_futures}, these have only evaluated one individual US region and/or have
only considered a small set of hours for their analysis. This paper is based
on data for the entire contiguous US of unprecedented temporal length and
spatial resolution. Relying on 32 years of weather data with hourly time
resolution and a spatial resolution of $40\times40$\,km$^2$, potential future
wind and solar PV generation time series are calculated and compared to
historical load profiles for the entire contiguous US, divided into the 10 FERC
(Federal Energy Regulatory Commission) regions (see Fig.\ \ref{fig:fercs}). 

We present two example applications of the obtained generation data: First,
it is examined how the mix of wind and solar power can be tuned to reduce the
usage of back-up power plants and storage, and second, different transmission
grid extensions and their effects are investigated. Both issues are first
addressed on a purely technical level, where our only concern is the reduction
of back-up or storage energy needed, and then on an economical level, taking
costs of wind and solar PV installations and of transmission lines into account.
These costs are resolved on a FERC region level to account for spatial
differences.

Comparisons between technically optimal systems and cost-optimal developments
allow us to judge the effect of costs as well as cost uncertainties on our
projections. For the mix between wind and solar PV power, we investigate
different relative costs in detail and show their impact on the optimal mix. For
transmission, we confine ourselves to two cost scenarios due to computational
limitations, and compare them to a heuristic approach used previously.

This paper is organized as follows: Sec.\ \ref{sec:2} describes the input weather
data set, and how wind and solar generation time series are obtained from it.
Sec.\ \ref{sec:3} describes the load data as well as the mismatch between VRES
generation and load. Secs.\ \ref{sec:4} and \ref{sec:5} present the two
applications of the generation and load data set: Calculation of the optimal mix
between wind and solar PV with respect to several objectives in Sec.\
\ref{sec:4}, and an optimal enhancement of the transmission grid for sharing
VRES among the FERC regions in Sec.\ \ref{sec:5}. Sec.\ \ref{sec:6} concludes
the paper.

\section{Weather and generation data}
\label{sec:2}

As data basis for the generation time series of wind and solar PV, we use
the renewable energy atlas developed in \cite{anders}. It is based on weather
data from NCEP CFSR for the years 1979-2010 \cite{saha}, with hourly time
resolution and an area of approximately 40$\times$40\,km$^2$ in each grid cell,
covering the contiguous US. The conversion from weather data to potential wind
and solar PV generation is done on a grid cell level and then aggregated to FERC
region level.

By aggregating time series of wind and solar power, we implicitly assume that
the FERC region-internal transmission system is essentially unconstrained. This
is reasonable for an electricity system with a high share of VRES, since
aggregation of wind and solar power smooths the total output \cite{archer,
holttinen05, sinden07, wiemken01, mills10, Widen:2011ys, Kempton:2010oq}, and
hence there is a strong incentive to remove bottlenecks in the transmission
grid. For now, we only assume aggregation on FERC level. The inter-FERC
transmission grid will be considered explicitly in Sec.\ \ref{sec:5}.

\subsection{Solar PV power}
\label{sec:2a}

\begin{figure}[!bh]
  \centering
  \includegraphics[width=0.49\textwidth]{\figdir/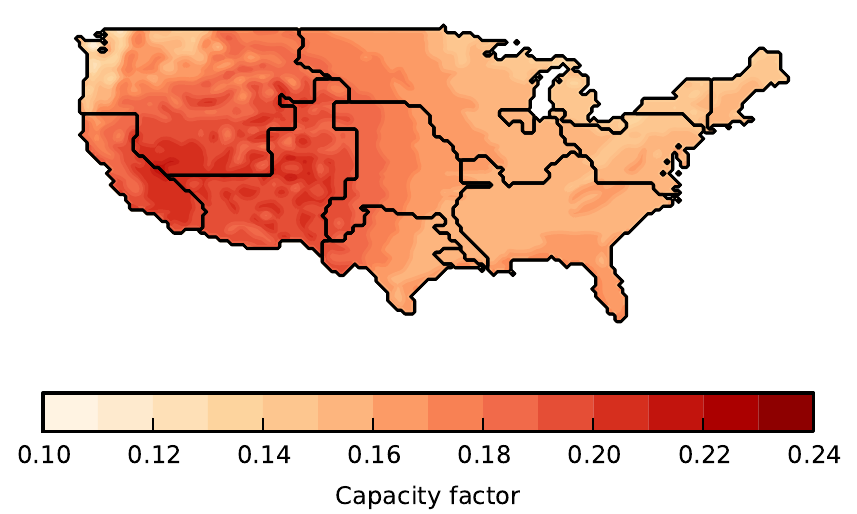}
  \caption{(Color online.) Solar resource map for the contiguous US as
    calculated from the renewable energy atlas \cite{anders}.
  } 
  \label{fig:pvres}
  \includegraphics[width=0.49\textwidth]{\figdir/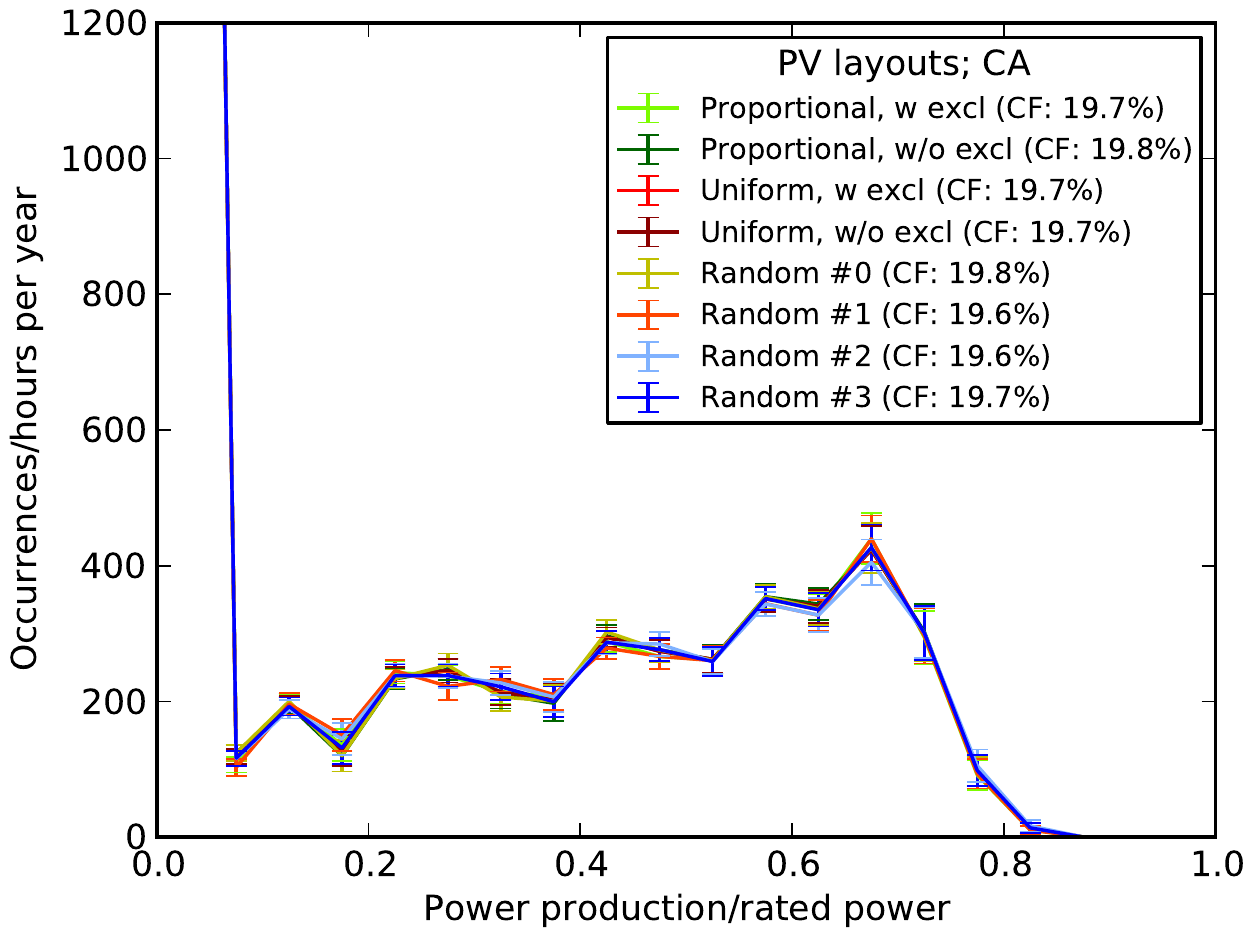}
  \caption{(Color online.) Solar power output histogram for California, for
    eight capacity layouts: Proportional to potential generation, not taking any
    excluded areas into account, proportional to potential generation, taking
    excluded areas from \cite{NREL_ewind,NREL_wwind} into account, uniform
    distribution with and without excluded areas, and four random layouts, in
    which solar power capacity is distributed randomly to 10\% of all grid
    cells. In the legend, the CF (capacity factor) of the layout that is
    achieved throughout the years is shown along with the layout name.
  }
  \label{fig:solval}
\end{figure}

Solar power production is calculated from weather data as detailed in
\cite{anders}, assuming non-tracking, south-oriented solar panels of the type
Scheuten~215~I \cite{scheuten215i} with a tilt equal to latitude. The
corresponding resource map is shown in Fig.\ \ref{fig:pvres}. It agrees very
well with the respective solar PV resource map from the National Renewable
Energy Laboratory (NREL) \cite{NREL_maps}.

The actual production within a FERC region is then determined by applying a
capacity layout to the grid cells, i.e.\ deciding how much capacity is installed
in each grid cell and summing up the output from all cells, weighted with this
layout. The validation plot Fig.\ \ref{fig:solval} shows the resulting
generation time series' production statistics for eight different capacity
layouts: Uniform distribution of PV capacity and distribution proportional to
the potential solar energy output both with or without exclusion of areas that
are declared unsuitable and/or prohibited according to \cite{NREL_ewind,
NREL_wwind}, and four layouts in which the PV capacity is assigned randomly to
10\% of the grid cells. The night hours amount to a peak a zero production. The
plot reveals that the choice of capacity layout does not have a large effect on
the (normalized) solar generation time series. The spread in capacity factor is
only 0.2\% for the example region of California shown in the plot. For FERC
regions in which the resource is less homogeneously distributed, such as ERCOT
or NW, a slightly larger spread of about 0.5\% is observed. To make a realistic
guess for the layout, we assume a capacity distribution proportional to the
potential of the grid cell under consideration with exclusion of unsuitable
areas. The solar capacity layout looks therefore very similar to the solar
potential map Fig.\ \ref{fig:pvres}.

\subsection{Wind power}
\label{sec:2b}

\begin{figure}[!tb]
  \centering
  \includegraphics[width=0.49\textwidth]{\figdir/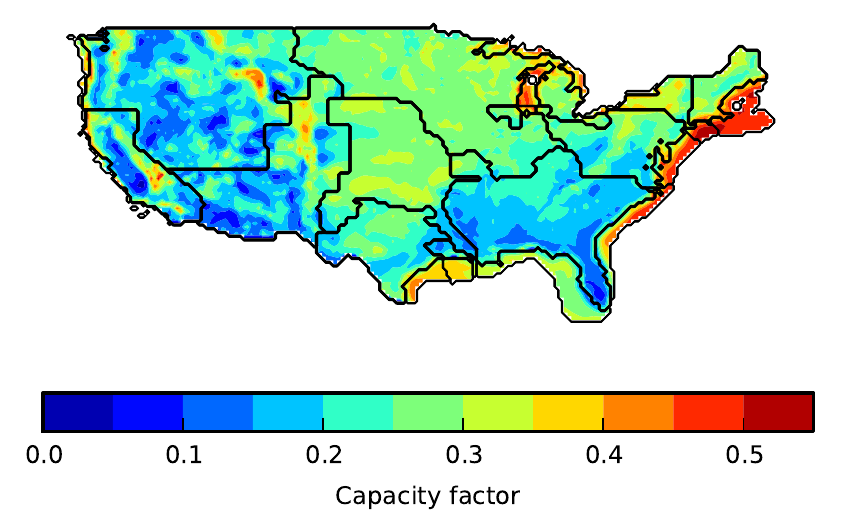}
  \caption{(Color online.) Wind resource map for the contiguous US as calculated
    from the renewable energy atlas \cite{anders}, modified as described in
    \ref{app:wind} to take effects of orography, surface roughness, and
    siting into account.
  } 
  \label{fig:wndres}
  \includegraphics[width=0.49\textwidth]{\figdir/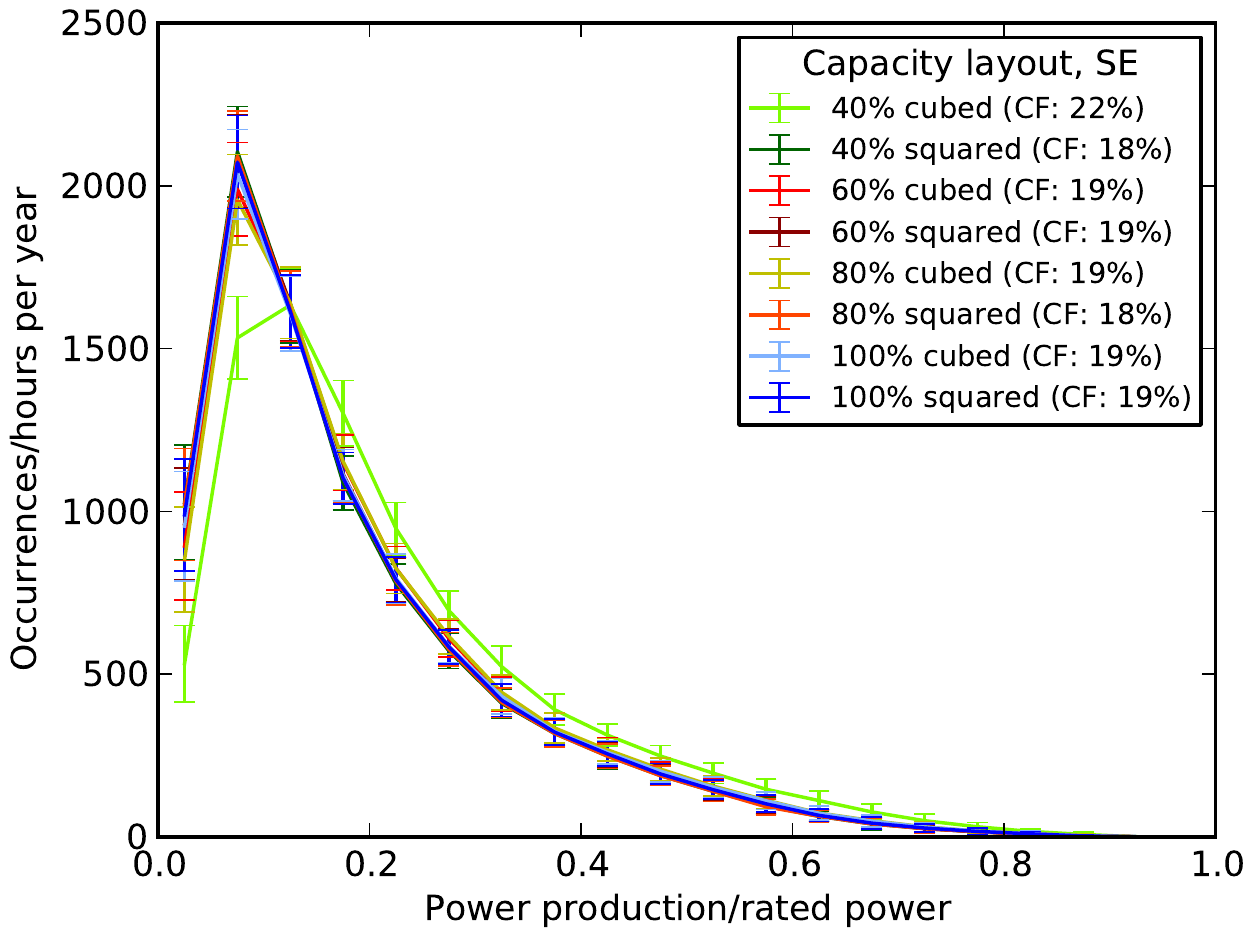}
  \caption{(Color online.) Wind power output distribution and capacity factors
    for different capacity layouts for the SE FERC region. The layouts are
    chosen randomly, with probability of picking a grid cell proportional to its
    wind potential squared or cubed as stated in the legend, and the total
    capacity was split into more or fewer units to be randomly distributed
    (percentage value in the legend), see Sec.\ \ref{sec:2b} for a detailed
    explanation.
  }
  \label{fig:wndval}
\end{figure}
\begin{figure}[!th]
  \centering
\includegraphics[width=0.49\textwidth]{\figdir/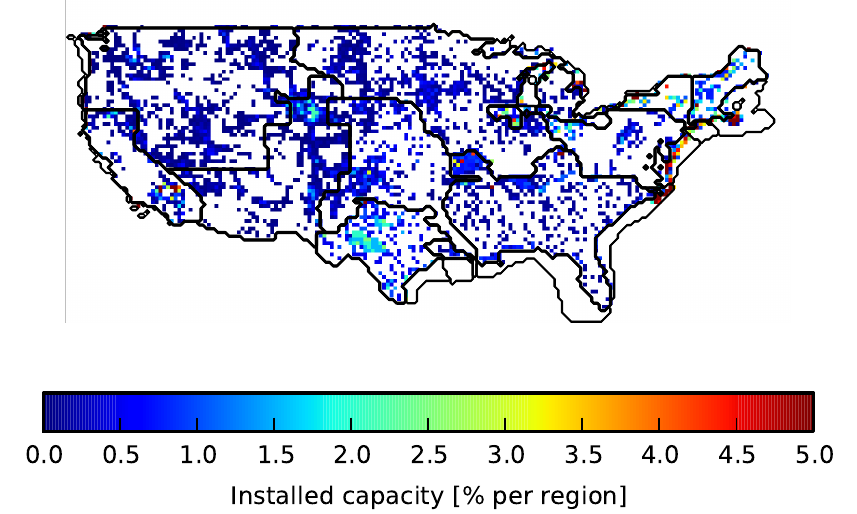}
  \caption{(Color online.) Wind capacity layout for the contiguous US used in
    this study. White cells do not contain any capacity. For the colored cells,
    color encodes the amount of capacity in percent of the installed capacity
    per FERC region. The 200\,m bathymetry line is shown offshore. For most FERC
    regions, a merger of the wind capacity layouts of \cite{NREL_ewind,
    NREL_wwind} is used. For SE and ERCOT, this is not possible due to a very
    low number of sites in these two FERC regions in both NREL datasets. We
    therefore use a synthetic random layout in SE and ERCOT, distributing a
    number of 40\% of the number of grid cells in the region of capacity units
    over all grid cells. The probability of picking a specific cell is chosen
    proportional to its potential power output cubed, and putting more than one
    capacity unit into a grid cell was allowed. See Sec.\ \ref{sec:2b} for a
    detailed explanation. This layout is seen to match the distribution of
    capacity across the rest of the US well.
  }
  \label{fig:cl}
\end{figure}
Wind speed interpolation from 10\,m wind data to hub height is used:
\begin{align}
  u(H) =  u(10{\rm m}) \frac{\ln\left(\frac{H}{z_0}\right)}
  {\ln\left(\frac{10{\rm m}}{z_0}\right)} \,,
  \label{eq:wextra}
\end{align}
where $H$ is the hub height, $z_0$ is the surface roughness, and $u$ is the wind
speed as a function of height. This vertical extrapolation tends to
underestimate hub wind speeds slightly, as discussed in \cite{archer03}. Their
research indicates that it would be better to use measurement data from
soundings. However, since such data are not available for the entire US, the
simple conversion method of Eq.\ \ref{eq:wextra} is employed. A hub height of
80\,m onshore and 100\,m offshore is chosen. To convert the wind speed at hub
height to power output, the power curve of the Vestas V90~3\,MW turbine is used
onshore, and the Vestas V164~7\,MW turbine offshore, as provided by the
manufacturer \cite{vestas}.  These relatively new and large models were chosen
since the main aim of this study is the investigation of a far future, highly
renewable energy system. The wind resource map thus obtained is shown in Fig.\
\ref{fig:wndres}, which aligns reasonably well with the resource maps from NREL
\cite{NREL_maps}. The conversion from wind speed data to wind power generation
was modified with the methods of \cite{BJ2010,BJ2011} to take effects of
orography, surface roughness, and siting into account, see \ref{app:wind} for
details. 

For wind, the sensitivity to siting is substantially higher than for PV, as is
observed from the spread in the production distribution for different capacity
layouts for wind (Fig.\ \ref{fig:wndval}), which is large compared to the
corresponding Fig.\ \ref{fig:solval}. We therefore rely on the wind capacity
layouts given by the Eastern and Western wind studies of NREL \cite{NREL_ewind,
NREL_wwind}, which include extensive siting analysis. Their layouts do not cover
the FERC regions ERCOT and SE very well. For these two regions, we use a
randomized layout. The wind power output distribution from eight different
candidate layouts for SE is shown in Fig.\ \ref{fig:wndval}, which compares
power output statistics. All of them are randomly generated by distributing a
number of capacity units across all available grid cells, proportional to their
potential wind power output squared or cubed (cf.\ the legend of Fig.\
\ref{fig:wndval}). The higher the exponent on the potential wind output, the
more high-yield sites are preferred. The amount of capacity units is a handle on
how smooth the layout becomes: The fewer units, the more grained the final
layout. It is chosen between 40\% and 100\% of the number of available grid
cells. Grid cells are allowed to hold more than one unit of capacity, so even in
a layout using 100\% of all grid cells as the number of capacity units, not all
grid cells are covered. Since the power output is normalized, only the relative
capacity fraction assigned to each grid cell is important. The layout picked for
SE and ERCOT in this analysis uses 40\% of all grid cells in capacity units,
distributed proportional to the cube of potential wind power output. It can be
seen in Fig.\ \ref{fig:cl} to match well the distribution of wind sites in the
rest of the US.

The mix between on- and offshore wind is chosen such that the relative capacity
between the two is the same as in the NREL wind studies \cite{NREL_ewind,
NREL_wwind}, see Tab.\ \ref{tab:onoffmix} for the values used. Wind
installations in the Great Lakes have been treated as offshore, i.e.\
the offshore 7\,MW turbine are assumed to be installed.
\begin{table}[!th]
  \centering
  \caption{Relative fraction of on- and offshore wind power installations for
    the layouts used in this study, for each FERC region separately.
  }
  \label{tab:onoffmix}
  \begin{tabular}{lrr}
    \hline
    Region & onshore fraction & offshore fraction \\
    \hline
    AllCA  &  98.2\% &   1.8\% \\
    ERCOT  & 100.0\% &   0.0\% \\
    ISONE  &  45.8\% &  54.2\% \\
    MISO   &  97.6\% &   2.4\% \\
    NW     &  99.9\% &   0.1\% \\
    NYISO  &  60.8\% &  39.2\% \\
    PJM    &  42.3\% &  57.7\% \\
    SE     & 100.0\% &   0.0\% \\
    SPP    & 100.0\% &   0.0\% \\
    SW     & 100.0\% &   0.0\% \\
    \hline \\
  \end{tabular}
\end{table}

\section{Load data and mismatch}
\label{sec:3}

Actual, historical (2006-2007) load data on the FERC region spatial scale with
hourly temporal resolution from the TSOs (transmission system operators) as
compiled in \cite{bethany} serves as the third ingredient to calculate the
hourly mismatch between VRES generation and load on FERC region level:
\begin{align}
  \Delta_n(t) = \gamma_n\left(\alpha_n^{\rm W} G_n^{\rm W}(t) 
                + (1-\alpha_n^{\rm W})G_n^{\rm S}(t)\right)
                \cdot \langle L_n \rangle
                - L_n(t)
  \label{eq:mism}
\end{align}
In this equation, $G_n^{\rm W}(t)$ and $G_n^{\rm S}(t)$ are the wind and solar
PV generation, respectively, in FERC region $n$, at time $t$, normalized to an
average of unity. $L_n(t)$ is the load in FERC region $n$, at time $t$, in MW,
and $\langle L_n \rangle$ is its time average. $\gamma_n$ is the renewable
penetration, i.e.\ the gross share of VRES. It is used as a scaling factor to
model different stages of the VRES deployment. Finally, $\alpha_n^{\rm W}$ is
the relative share of wind in VRES.

The load data are extended by repetition to cover the entire timespan of wind
and solar data of 32 years. To this end, the load of the SW FERC region was
de-trended by removing a net linear growth such that the end of 2007 and the
beginning of 2006 fit together. For all other FERC regions, this was not
necessary.

\section{Optimal mixes}
\label{sec:4}

As a first example application of the obtained US wind and solar generation
data, we look at three different ways of optimizing the mix between wind and
solar PV: minimizing storage energy capacity, minimizing system imbalance
energy, and minimizing levelized costs of renewable electricity generation. The
mixes are all calculated for a fully renewable scenario, i.e.\ a VRES gross
share of 100\%, in a scenario where each FERC region operates independently
(that is, no inter-FERC transmission) as well as for full aggregation across the
entire US (corresponding to unlimited transmission). First, we minimize storage
energy capacity in the case of no other sources of back-up energy. Then, we
minimize system imbalance in the case where this imbalance is provided by
weather-independent dispatchable generators without storage. Finally, we
minimize wind and solar levelized cost of electricity (LCOE), taking the
regionally-adjusted costs of wind and solar PV into account, again with
dispatchable back-up generation and without storage. Collectively, these three
cases give us insight into how the optimal wind and solar mix for a fully
renewable US electric system varies for different system criteria, and what
benefits are to be gained by adjusting the wind/solar mix.

\subsection{Minimizing storage energy capacity}
\label{sec:4a}

A scenario is considered where each of the FERC regions is isolated from the
others and each of them have reached a VRES gross share of 100\%,
$\gamma_n=1\,\forall n$ in Eq.\ \eqref{eq:mism}. For comparison, the analogous
results for an aggregation across the entire US is calculated as well. We look
at an electricity system where all the surplus generation (positive mismatch in
Eq.\ \eqref{eq:mism}) is put into an idealized, 100\% efficient storage system
and all deficits are covered by re-extracting the stored energy. Since VRES
generation equals on average the load and storage losses are neglected, such a
system provides enough power at all times.

Our objective is to minimize the storage energy capacity $E_n^H$. It can be
calculated from the storage filling level time series $H_n(t)$ as follows:
\begin{align}
  H_n(t) &= H_n(t-1) + \Delta_n(t) \nonumber \\
   E_n^H &= \max_t (H_n(t)) - \min_t (H_n(t))
  \label{eq:minstor}
\end{align}
The storage optimal mix for FERC region $n$ is defined to be the $\alpha_n^{\rm
W}$ that minimizes this quantity.
\begin{figure}[!bt]
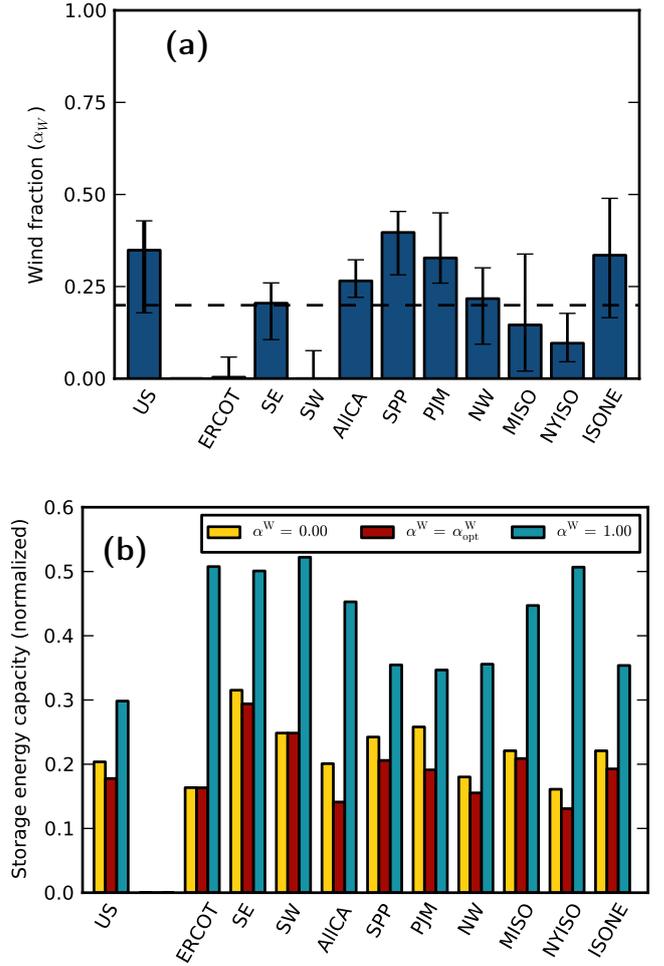

  \centering
  \begin{lpic}{\figdir/optimal_mix_storage_US_p_interval_1.0(3.5in)}
    \lbl{25,56;{\large \textsf{\textbf{(a)}}}}
  \end{lpic}
  \begin{lpic}{\figdir/storage_energy_capacity_vs_mix_US(3.5in)}
    \lbl{17,56;{\large \textsf{\textbf{(b)}}}}
  \end{lpic}
  \caption{(Color online.) (a): Storage optimal mix between wind and solar
    power, given as the percentage of wind power, for the contiguous US FERC
    regions as well as their aggregation (marked "US"), at 100\% renewable
    penetration. This mix leads to minimal storage energy capacity needs
    (assuming that all residual loads have to be covered from a stored surplus;
    no storage losses). The error bars indicate mixes that lead to a storage
    energy capacity that is larger by one percent of the load than for the
    storage optimal mix. The dashed line marks the weighted average of the
    storage optimal mixes across all FERC regions. (b): Storage energy capacity,
    normalized by the average annual load, for different mixes of wind and solar
    PV power: Solar PV only, the storage optimal mix, and wind only.
  }
  \label{fig:optmix_sto}
\end{figure}

The mix minimizing storage energy capacity needs is heavily leaning toward solar
PV power, leading to almost exclusive use of solar for the southernmost FERC
regions, see Fig.\ \ref{fig:optmix_sto}a. This is due to the general trend that
solar irradiation shows less seasonal variation close to the equator, and is
therefore more favorable in terms of storage needs, since these are mainly
determined by seasonal timescales \cite{heide2011}. Additionally, the load in
most of the US peaks in summer due to air conditioning needs. It is thus
correlated with the solar PV power output, further shifting the US storage
optimal mix toward solar PV. This mix may change when the seasonal load pattern
in the US changes, which may happen, e.g.\ due to more electrical vehicles being
used and needing to be charged throughout the year. In contrast to the US, wind
gains a higher share in the European storage optimal mixes, which are on the
order of 50\%-60\% wind power \cite{heide2011}. This is due to two effects: The
load in Europe peaks in winter due to heating and illumination needs and is thus
anti-correlated to solar PV, and because of the higher latitudes, the seasonal
variation in solar PV output is more pronounced. The aggregation of the entire
contiguous US favors a higher share of wind, as shown in the leftmost bar of
Fig.\ \ref{fig:optmix_sto}a.

From Fig.\ \ref{fig:optmix_sto}a, it is apparent that the sensitivity of the
storage energy capacity to the mix is not very pronounced: A large change in the
mix leads to a rather small change in storage energy capacity. The error bars in
Fig.\ \ref{fig:optmix_sto}a indicate mixes that lead to storage energy
capacities larger than the optimum by one percent of the load. They spread
across 10\% to 25\% relative share.

The optimal storage energy capacity shown in Fig.\ \ref{fig:optmix_sto}b is
around two to three months of average load, which is comparable with European
values \cite{heide2011}. This figure also shows that a wind-only power system
has a highly unfavorable effect on storage capacities needed, roughly tripling
the storage needs in extreme cases such as ERCOT and, interestingly, also NYISO,
which has very good wind resource quality, cf.\ Fig.\ \ref{fig:wndres}.

\subsection{Minimizing balancing energy}
\label{sec:4b}

In this case, no storage is assumed to be in place. Instead, whenever the
mismatch \eqref{eq:mism} is negative, the residual load has to be covered from
other, dispatchable power sources. We collectively term these "balancing", $B$,
since they balance the electricity system. 

This scenario prompts a different plausible objective to determine the optimal
mix $\alpha_n^{\rm W}$ of wind and solar power: Minimize the total amount of
balancing energy $\sum_{t} B_n(t)$:
\begin{align}
  \sum_{t}& B_n(t) = 
  \sum_{t} \left( \Delta_n(t)\right)_- 
  \label{eq:balmin} \\
  = \sum_{t}& \left[\gamma_n\left(\alpha_n^{\rm W}G_n^{\rm W}(t)
    + (1-\alpha_n^{\rm W})G_n^{\rm S}(t)\right)\langle L_n\rangle
    - L_n(t)\right]_- \nonumber
\end{align}
as a function of $\alpha_n^{\rm W}$, where $[x]_- = \max\{-x,0\}$ denotes the
negative part of a quantity $x$.

\begin{figure}[!hbt]
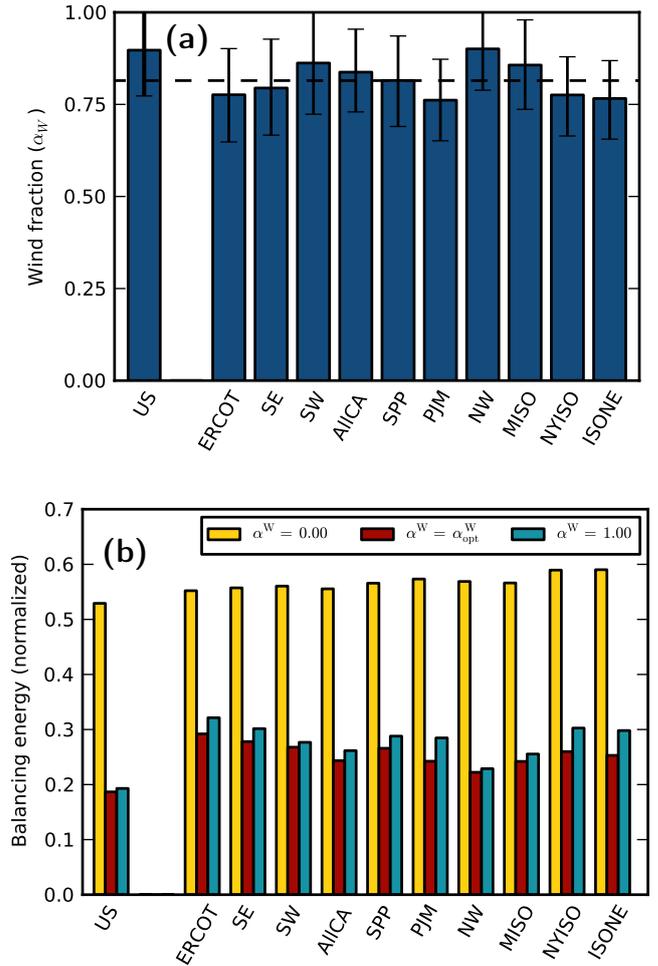

  \centering
  \begin{lpic}{\figdir/optimal_mix_balancing_US_p_interval_1.0(3.5in)}
    \lbl{25,57;{\large \textsf{\textbf{(a)}}}}
  \end{lpic}
  \begin{lpic}{\figdir/balancing_energy_vs_mix_US_CS_0(3.5in)}
    \lbl{17,56;{\large \textsf{\textbf{(b)}}}}
  \end{lpic}
  \caption{(Color online.) (a): Balancing optimal mix between wind and solar
    power, given as the percentage of wind power, for the contiguous US FERC
    regions, at 100\% renewable penetration. This mix minimizes balancing energy
    or equivalently, residual load. The error bars extend to mixes that would
    lead to one additional percent of the total load being covered from
    dispatchable sources. The dashed line indicates the weighted average of the
    balancing optimal mix across all FERC regions. (b): Balancing energies for
    different mixes: Solar PV only, the balancing optimal mix, and wind only.
  }
  \label{fig:optmix}
\end{figure}

In Fig.\ \ref{fig:optmix}a, the balancing optimal mix between wind and solar
power is shown for the contiguous US FERC regions. It is seen that the mix
minimizing residual load is around 80\% wind, almost homogeneously distributed
throughout the country. Again, for the aggregated contiguous US, the share of
wind is seen to rise, in this case to 90\%. This is due to long-range
decorrelation effects in the range of 500~km to 1000~km \cite{Widen:2011ys,
Kempton:2010oq}. The total balancing energy necessary is shown in Fig.\
\ref{fig:optmix}b. Single-region values range from a little less than 25\% to
about 30\% of the annual load. In the case of balancing energy minimization, the
solar PV-only mix is found to perform worst, which is due to the need for
balancing whenever the sun does not shine, that is, every night. Both the
balancing optimal mix and the optimal balancing energies are similar to what has
been calculated earlier with the same method for Europe \cite{rolando}. The only
noticeable deviation occurs in the fully aggregated case, where for Europe
optimal balancing energies as low as 15\% of the annual load have been found,
compared to 18\% for the contiguous US, see also Sec.\ \ref{sec:5a}. 

\subsection{Minimizing LCOE from VRES}
\label{sec:4c}

We now calculate an optimized mix of wind and solar PV power based on their LCOE
(levelized costs of electricity). As in Sec.\ \ref{sec:4b}, the contiguous US
with 100\% gross share of VRES are considered ($\gamma_n = 1$). It is assumed
that no storage system is in place. Surplus generation (positive $\Delta_n(t)$
in Eq.\ \eqref{eq:mism}) is curtailed, while insufficient generation has to be
balanced by dispatchable power. 

\subsubsection{Regional LCOE}
\label{sec:43a}

\begin{table}[!th]
  \centering
  \caption{Regional LCOE with corresponding regional capacity and weight
    factors, as well as the deviation from the average LCOE in percent, for wind
    (top) and solar (bottom). The US average LCOE for both wind and solar PV is
    assumed to be 0.08\,\$/kWh. Note that the NW FERC region also comprises
    Nevada and Utah (cf.\ Fig.\ \ref{fig:fercs}), thus explaining the low solar
    LCOE there.
  }
  \label{tab:LCOE}
  \begin{tabular}{lrrrrrr}
    \hline
    Region & $CF_n$ & $w_n$ & $m_n$ & dev.\ from avg. &
    $\frac{LCOE_n}{\text{\$/kWh}}$ \\
    \hline
    AllCA & 0.24 & 1.13 & 1.04 & 17\% & 0.094 \\
    ERCOT & 0.22 & 1.24 & 0.97 & 20\% & 0.096 \\
    ISONE & 0.39 & 0.71 & 1.02 & -28\% & 0.058 \\
     MISO & 0.28 & 0.98 & 1.00 & -2\% & 0.079 \\
       NW & 0.24 & 1.16 & 1.00 & 16\% & 0.093 \\
    NYISO & 0.35 & 0.79 & 1.04 & -17\% & 0.066 \\
      PJM & 0.34 & 0.81 & 1.01 & -18\% & 0.065 \\
       SE & 0.22 & 1.26 & 0.98 & 23\% & 0.099 \\
      SPP & 0.29 & 0.95 & 0.98 & -7\% & 0.074 \\
       SW & 0.28 & 0.97 & 0.99 & -4\% & 0.077 \\
     avg. & 0.29 & 1.00 & 1.00 & 0\% & 0.080 \\
    \hline
  \end{tabular}
  \begin{tabular}{lrrrrrr}
    \hline
    Region & $CF_n$ & $w_n$ & $m_n$ & dev.\ from avg. &
    $\frac{LCOE_n}{\text{\$/kWh}}$ \\
    \hline
    AllCA & 0.16 & 0.80 & 1.04 & -17\% & 0.066 \\
    ERCOT & 0.14 & 0.92 & 0.97 & -11\% & 0.071 \\
    ISONE & 0.11 & 1.22 & 1.02 & 24\% & 0.100 \\
     MISO & 0.12 & 1.06 & 1.01 & 7\% & 0.085 \\
       NW & 0.15 & 0.88 & 1.00 & -12\% & 0.071 \\
    NYISO & 0.11 & 1.22 & 1.10 & 35\% & 0.108 \\
      PJM & 0.11 & 1.13 & 1.03 & 16\% & 0.093 \\
       SE & 0.13 & 0.99 & 0.94 & -7\% & 0.074 \\
      SPP & 0.14 & 0.95 & 0.96 & -9\% & 0.073 \\
       SW & 0.15 & 0.84 & 0.98 & -18\% & 0.066 \\
     avg. & 0.13 & 1.00 & 1.00 & 0\% & 0.081 \\
    \hline
  \end{tabular}
\end{table}
The storage and balancing optimal mixes discussed above are based solely on the
temporal characteristics of the wind and solar generation. They do not take the
capacity factor into account, i.e.\ the ratio of the average generation of a
solar panel or wind turbine under the conditions at a given site to its
nameplate capacity. This number, however, determines how large an installation
needs to be in order to generate a certain amount of electrical energy, and is
consequently a major constituent of the total energy costs. Furthermore, the cost
of labor, materials and equipment varies locally across the US, leading to extra
regional differences in investment costs. We have developed a model to
incorporate these effects. Additionally, it is able to handle different US
average wind and solar LCOE.

First, regional LCOE variations due to capacity factor differences are included by
using the inverse capacity factors $CF_n$ as weights $w_n$. The costs of an
installation per unit of power are largely independent of the number of units of
energy generated. Since the cost is distributed evenly over all units of energy
generated, the cost of a single unit is directly anti-proportional to the number
of units generated. The weights are normalized to keep the average price at
input level.
\begin{align*}
  \textstyle w_n &= \frac{CF_n^{-1}}{\sum_m CF_m^{-1}/10} \\
\end{align*}

Second, different labor, equipment, and materials costs in different FERC regions
are taken into account using the method described in \cite{E3report} with input
data from \cite{cwccis}. This yields regional multipliers $m_n$ on the order of
one for each of the FERC regions. Put together, this results in a local LCOE of
\begin{align}
  \text{LCOE}_n &= m_n w_n \cdot \text{LCOE}_\text{avg}\,,
  \label{eq:reg}
\end{align}
where $\text{LCOE}_\text{avg}$ is the average LCOE of the technology under
consideration.  This cost regionalization is done twice independently, once for
wind and once for solar. The cost factors as well as the regional LCOE can be
found in Tab.\ \ref{tab:LCOE}. It is observed that solar installations have
lowest costs in the southern and western regions, while they are expensive on
the northern East Coast, and vice versa for wind costs.

\subsubsection{Calculation of the LCOE-optimal mix}
\label{sec:43b}

The regional LCOE of wind and solar are then combined and modified to include
the effects of curtailment by multiplying them by the ratio of generated to used
energy:
\begin{align}
  &\text{LCOE}_0(\alpha_n^{\rm W}) = \alpha_n^{\rm W} \text{LCOE}_n^{\rm W} +
  (1-\alpha_n^{\rm W})\text{LCOE}_n^{\rm S} \\
  &\text{LCOE}_\text{mod.}(\alpha_n^{\rm W}) =  
  \phantom{\frac{E_i}{E_i}}\nonumber \\
  &\qquad \text{LCOE}_{0}(\alpha_n^{\rm W})\cdot
  \frac{E_\text{generated}(\alpha_n^{\rm W})}
  {E_\text{generated}(\alpha_n^{\rm W})-E_\text{curtailed}(\alpha_n^{\rm W})}
  \label{eq:prices1}
\end{align}
This reflects that the LCOE are incurred for all the energy generated, but only
recovered by sales of the non-curtailed part (in an idealized economy where
retail prices equal the LCOE). Surplus generation thus becomes undesirable in
this formulation, because it leads to an effective rise in LCOE.

\subsubsection{LCOE-optimal mix}

If the LCOE of wind and solar are equal in a given FERC region, the LCOE optimal
mix reduces to the balancing minimization discussed in Sec.\ \ref{sec:4b}. In
this case, LCOE$_0$ in Eq.\ \eqref{eq:prices1} becomes independent of the mix,
and since the total generated VRES energy $E_\text{generated}$ is constant, the
optimum is found when $E_\text{curtailed}$ is minimal. Since the average VRES
generation equals the average load, the total curtailed energy is equal to the
\begin{figure}[!bh]
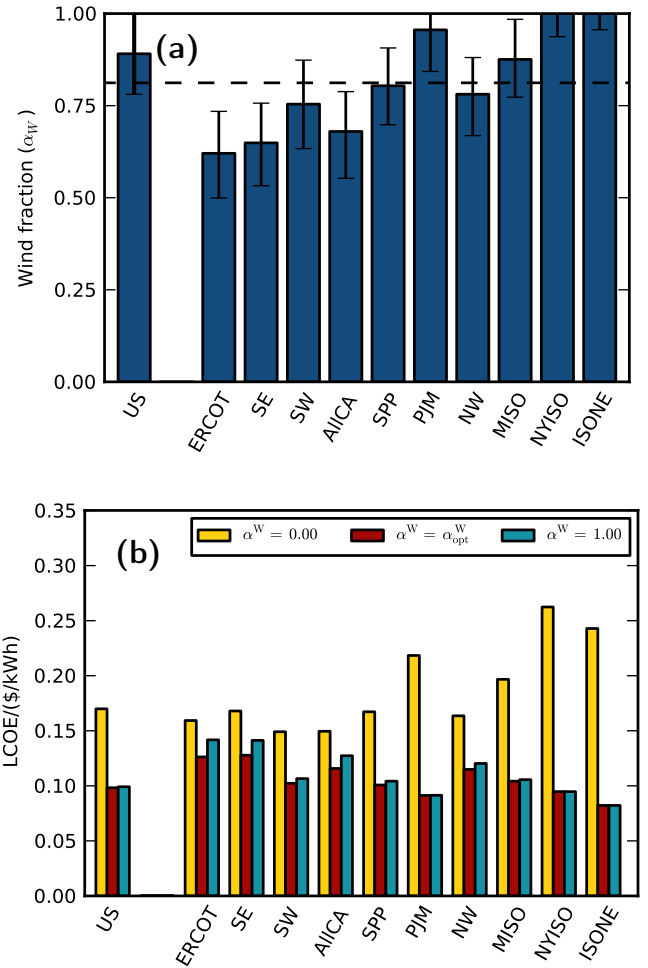

  \centering
  \begin{lpic}{\figdir/optimal_mix_balancing_US_p_interval_1.0_LCOE_based_wind_0.080_solar_0.080(3.5in)}
    \lbl{25,56;{\large \textsf{\textbf{(a)}}}}
  \end{lpic}
  \begin{lpic}{\figdir/all_LCOE_vs_mix_US_LCOE_wind_avg_0.080_LCOE_solar_avg_0.080(3.5in)}
    \lbl{20,56;{\large \textsf{\textbf{(b)}}}}
  \end{lpic}
  \caption{(Color online.) (a) LCOE-optimal mixes for the case of equal average
    wind and solar LCOE of 0.08\,\$/kWh, calculated as described in Secs.\
    \ref{sec:43a} and \ref{sec:43b}. Note that (a) does not reproduce Fig.\
    \ref{fig:optmix}, because although the mean LCOE for wind and solar are the
    same here, they are not the same for all FERC regions. (b) Corresponding
    LCOE in the FERC regions as well as on an aggregated level (denoted "US"),
    of different wind and solar mixes: Solar PV only, LCOE optimal mix, and wind
    only.
  }
  \label{fig:cost_mix}
\end{figure}
total balancing energy, and therefore minimal curtailment and minimal balancing
are equivalent here.

We first investigate a case where wind and solar power have the same average
LCOE of 0.08\,\$/kWh (before taking the regionalization from Eq.\ \eqref{eq:reg}
or the curtailment corrections from Eq.\ \eqref{eq:prices1} into account). Due
to the regionalization of LCOE, this translates into different LCOE in the
different FERC regions, and thus we do not simply reproduce the results of Sec.\
\ref{sec:4b}. Comparing the regional LCOE optimal mixes shown in Fig.\
\ref{fig:cost_mix}a to the balancing optimal mixes in Fig.\ \ref{fig:optmix}a,
we see a shift of the mix toward wind in the North East (particularly in ISONE
and NYISO), where wind resources are very good and local wind LCOE are thus low,
while it is shifted toward solar in the South (particularly in ERCOT, SE and
AllCA) because of the good solar resources there.

The LCOE of different mixes (LCOE-optimal, solar only and wind only) is shown in
Fig.\ \ref{fig:cost_mix}b. It is apparent that picking the LCOE-optimal mix is
able to reduce average LCOE significantly, especially compared to solar-only
scenarios. For example, for the entire US, solar only is 70\% more expensive
than the optimal mix, and for the North East, it is more than twice as
expensive. The LCOE for the aggregated US are about \$0.10, as could have been
directly predicted from Eq.\ \eqref{eq:prices1}: Since wind and solar LCOE are
equal in this case, the LCOE optimal mix equals the balancing optimal mix. As
calculated in Sec.\ \ref{sec:4b}, the optimal balancing energy, which equals the
curtailed energy, is 18\% of the load. Eq.\ \eqref{eq:prices1} thus yields
\[
  \text{LCOE}_\text{mod.} = \text{LCOE}_0 \cdot \frac{1}{1-0.18} 
  = \frac{\$0.08}{0.82} \approx \$0.10\,.
\]
\begin{table*}[!bt]
  \centering
  \caption{Levelized cost of electricity (LCOE) from various reports, as
    compiled by Open Energy Information (OpenEI \cite{openei}). For wind, only
    two reports were available, so the average is used instead of the median,
    and quartiles are not meaningful and therefore omitted. In order to
    calculate a mean price for wind (on- and offshore), a mix of 25\% offshore
    and 75\% onshore installations is assumed.
  }
  \label{tab:prices}
  \begin{tabular}{lrrrrrr}
    \hline
    Technology  & \multicolumn{5}{c}{LCOE in \$/kWh}    & \# of
    reports \\
    & minimum & 1st quartile & median & 3rd quartile & maximum & \\
    \hline
    Wind (onshore) & 0.060 & - & 0.065 & - & 0.070 & 2 \\
    Wind (offshore) & 0.100 & - & 0.105 & - & 0.110 & 2 \\
    Wind (75/25 mix) & 0.070 & - & 0.075 & - & 0.080 & 2 \\
    Solar PV & 0.040 & 0.080 & 0.120 & 0.190 & 0.240 & 12 \\
    \hline
  \end{tabular}
\end{table*}

\subsubsection{Sensitivity to different average LCOE ratios}

Today, wind and solar PV differ significantly in their installation price, and
they may continue to do so in the future. Various projections of average LCOE
for wind and solar PV power across the US have been compiled by Open Energy
Information (OpenEI) \citep{openei}. We use the price projections for 2020 from
the most recent available reports (from 2012) to illustrate the large LCOE
ranges, see Tab.\ \ref{tab:prices}.

\begin{figure}[!bt]
  \centering
  \includegraphics[width=0.49\textwidth]{\figdir/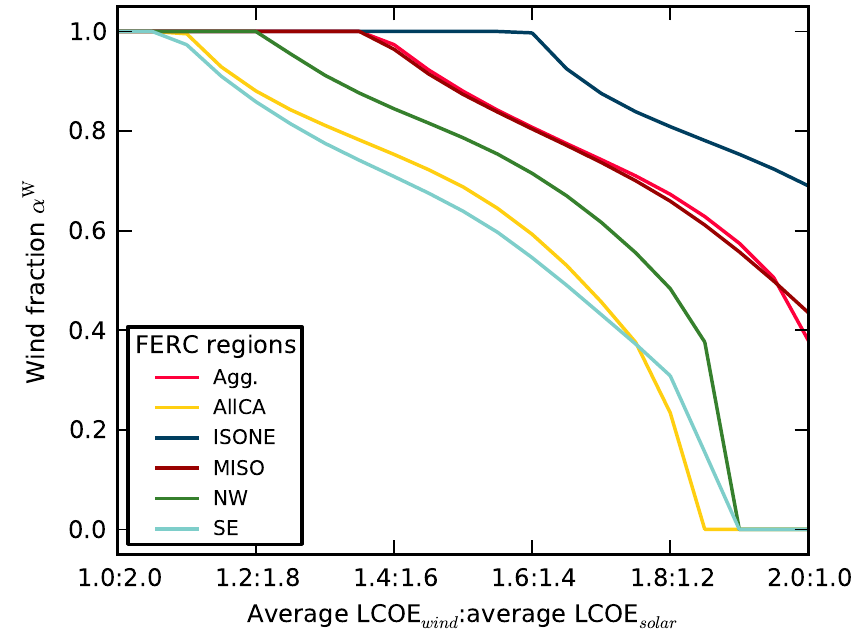}
  \caption{(Color online.) Cost-optimal wind/solar mixes for five geographically
    representative FERC regions, for different ratios of the LCOE of wind and
    solar power. The regional differences in mix due to the climate differences
    are clearly visible: Southern FERC regions such as AllCA and SE switch to
    100\% solar if this is much cheaper than wind, while other FERC regions
    never do. Conversely, if wind is much cheaper, all FERC regions switch to
    100\% wind power, starting from the north eastern regions, here represented
    by ISONE.
  } 
  \label{fig:optmix_sweep}
\end{figure}

The impact of different wind and solar price ratios is depicted in Fig.\
\ref{fig:optmix_sweep}, which shows the regional LCOE-optimal mix as a function
of the ratio of average LCOE, for the representative FERC regions of AllCA,
ISONE, MISO, NW, and SE, as well as for the aggregated US. We see that if wind
LCOE are half of the solar LCOE on average, then 100\% wind will be the optimal
mix for all FERC regions. Conversely, if average solar LCOE are half of the
wind LCOE, this does \emph{not} lead to a 100\% solar cost optimal mix in all
FERC regions, cf.\ Fig.\ \ref{fig:optmix_sweep}. This is due to the large
curtailment and balancing such a mix entails. As seen in Sec.\ \ref{sec:4b}, the
mix minimizing balancing energy lies around 80\% wind. Solar only is very
unfavorable because it would lead to large balancing and curtailment needs, even
if solar LCOE were much lower than wind LCOE. 

Looking at different LCOE combinations, we observe that the results in Fig.\
\ref{fig:cost_mix}a (i.e.\ when solar and wind have the same cost) most closely
match those from the NREL Renewable Electricity Futures study
\cite{NREL_re_futures}, which uses a cost-based optimization tool to determine
the least-cost portfolio of generators, storage, and transmission for various
scenarios of an 80\% renewable US electric system. In the NREL study, for the
2050 LCOE values for comparable installed capacity values, wind is more heavily
installed in the Great Plains, Great Lakes, Central, Northwest, and Mid-Atlantic
areas (roughly corresponding to the MISO, SPP, NW and PJM FERC regions) and
solar is more heavily installed in CA, the Southwest, Texas, and the South
(roughly corresponding to AllCA, SW, ERCOT, and SE FERC regions).

The high sensitivity of solar PV to prices, relative to that of wind, is
corroborated by the significant price impact on solar PV build-out observed in
\cite{Nelson12}. The NREL Renewable Electricity Futures Study also recognized a
high sensitivity of the solar energy (PV and CSP) build-out to varying cost
estimates \cite{NREL_re_futures}. Furthermore, they found that the relative
contributions of wind and solar generation were on average 75\% wind to 25\%
solar across the contiguous US in their optimized scenario. This agrees well
with the cost-optimal $\alpha^{\rm W}$ value for equal wind and solar LCOE found
here of slightly less than 80\%.

\section{Optimal transmission grid extensions}
\label{sec:5}

After studying different mixes of wind and solar PV power by optimizing various
objectives, we now set out to investigate a further component of the electricity
system: The role of transmission. In this section, we use the same setting as
presented in Sec.\ \ref{sec:4b}, that is, the mix between wind and solar PV
power is kept at the single node balancing optimal mix, gross VRES share is
again 100\%, and no storage system is included, such that all deficits have to
be covered by balancing. 

We first make some general observations and introduce the generalized DC power
flow to be used, and then examine three different transmission layouts. Their
size is chosen such that they can all be built for a capital investment twice as
large as the cost of existing installations. The goal set for transmission grid
extensions is that they should reduce balancing energy usage as much as
possible. It is not a priori clear how the given investment should be
distributed to reinforce single lines in order to achieve this, so we develop
different methods of assigning capacity to single links. The first layout is
done based on quantiles of the distribution of unconstrained flows analogous to
the studies in Refs.\ \cite{rolando,sarah} for Europe ("Quantile layout"). The
second is done by cost optimization for the hypothetical case where all lines
have the same costs ("Even layout"). The last one uses cost optimization
adopting realistic line cost estimates from Refs.\ \cite{reeds, bethany} ("Real
layout"). The latter two layouts are obtained by the optimization technique of
simulated annealing. Our implementation of the algorithm is introduced before
the optimized layouts are discussed.

\subsection{Maximal balancing energy reduction}
\label{sec:5a}

The first thing to observe is that the maximal possible balancing energy
reduction from transmission can be calculated ad hoc, just by comparing the
isolated balancing needs,
\begin{align}
  B_\text{tot}^\text{isolated} = 
  \sum_{t} \sum_{n} \left[\Delta_n(t)\right]_-\, ,
  \label{eq:iso}
\end{align}
with the aggregated ones,
\begin{align}
  B_\text{tot}^\text{aggregated} = 
  \sum_{t} \left[\sum_{n} \Delta_n(t)\right]_-\, .
  \label{eq:agg}
\end{align}
In Eq.\ \eqref{eq:iso}, all negative mismatches for the different nodes are
summed up, yielding the total balancing energy in the case of isolated nodes.
Meanwhile, in Eq.\ \eqref{eq:agg} the mismatches are first added, thus allowing
a negative mismatch at one node to be canceled by a positive one at another.
The negative part of this aggregated mismatch, summed over all time steps, gives
the minimal possible amount of balancing energy. For the contiguous US, these
two numbers are $B_\text{tot}^\text{isolated} = 25.7\%$ of the total load
covered from balancing energy for the isolated case, compared to
$B_\text{tot}^\text{aggregated} = 19.0\%$ in the aggregated case when keeping
the wind/solar mixes fixed (no optimization of the mix for the aggregated US as
in Sec.\ \ref{sec:4b}). Transmission can thus effect a balancing energy
reduction by roughly a quarter. Compared to the corresponding scenario for
Europe, the isolated nodes (countries in the European case) have to balance
around 24\% of the total load, which drops to around 15\% in the aggregated
case, thus a reduction by about two fifths \cite{rolando}. This indicates that
although Europe covers a smaller area, low production phases of wind and solar
PV are less correlated there, and hence the aggregated output is smoother than
for the US.

\subsection{Generalized DC power flow}
\label{sec:5b}

The flow paradigm introduced and described in Refs.\ \cite{rolando,sarah} is
used to calculate the distribution of balancing as well as the flows on the
single links. In this formulation, the standard DC power flow, which is a valid
approximation for the full AC flow under stable grid conditions, is generalized
to cope with flow capacity constraints and global mismatches. 

The directed flow along link $l$ is denoted $F_l$. It is constrained by
(possibly direction dependent) power flow capacities $h_{\pm l}$ of the link
$l$, 
\[
  h_{-l}\leq F_l \leq h_l\,.
\]
Furthermore, we make use of the incidence matrix $K$ which encodes the network
topology:
\[
  K_{nl} = \begin{cases}  
    \phantom{-}1 & \text{if link $l$ starts at node $n$}\\
              -1 & \text{if link $l$ ends at node $n$}\\
    \phantom{-}0 & \text{else}
  \end{cases}
\]
Start and end point of each link can be chosen arbitrarily, they only have to be
used consistently throughout the calculations. With the help of the flow vector
and the incidence matrix, the net outflow from node $n$ can be expressed as
\[
  \sum_l K_{nl} F_l\,.
\]
If this quantity is negative, the node experiences a net inflow. The goal is now
to find a flow vector $(F_l)_{l=1..L}$ that leads to imports and exports at the
single nodes such that deficits and excesses are canceled out at all the nodes,
while observing the flow constraints. Since there is in general a either a
global excess or a global deficit in the grid and the total energy is conserved,
it is not possible to reduce all deficits and excesses to zero. Instead, use the
following procedure:
\begin{align}
  & \min_{h_{-l}\leq F_l \leq h_l} B_\text{tot}
  = \min_{h_{-l}\leq F_l \leq h_l} \sum_n \left[ \Delta_n - (KF)_n\right]_- 
  = B_\text{min} 
  \label{eq:step1} \\
  & \min_{\substack{h_{-l}\leq F_l \leq h_l \\ \sum_{i=1}^N (\Delta_n-(K\cdot F)_n)_- = B_{\rm min}}} \sum_l F_l^2
  \label{eq:step2}
\end{align}
In the first step, Eq.\ \eqref{eq:step1}, the sum of all deficits after imports
and exports is minimized. This corresponds to using as little balancing energy
or equivalently as much VRE as possible. In the second step, Eq.\
\eqref{eq:step2}, flow dissipation is minimized, which is proportional to the
sum of all the flows squared, while keeping the total deficit at its minimal
value found in the first step. This algorithm entails that excesses and deficits
at the nodes are matched as locally as possible. For example, if there is a
deficit at node A, it is preferred to import to A from nodes in A's neighborhood
instead of farther away nodes.

\subsection{Quantile capacity layouts}
\label{sec:5c}

\begin{figure}[!tbh]
  \centering
  \includegraphics[width=0.49\textwidth]{\figdir/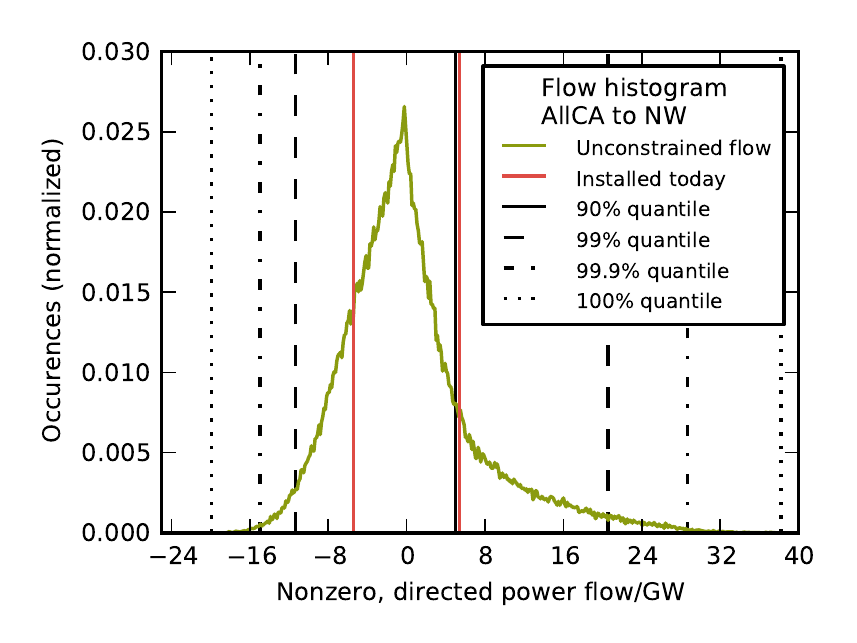}
  \caption{(Color online.) Unconstrained, directed flow distribution on the
    example link between AllCA and NW. Vertical lines indicate today's
    installation (red line), 90\% (solid line), 99\% (dashed line), 99.9\%
    (dashed-dotted line), and 100\% (dotted line) quantiles. Note that in the
    negative direction (from NW to AllCA), the solid black 90\% quantile line is
    almost covered by today's installation (red vertical line).
  }
  \label{fig:quant}
\end{figure}

Neglecting different costs for different lines, the best grid build-up found so
far (to our knowledge) is what we term "Quantile line capacities"
\cite{rolando}. These are calculated by first solving the power flow, Eqs.\
\eqref{eq:step1} and \eqref{eq:step2}, without the constraint $h_{-l}\leq
F_l\leq h_l$, for all hours in the time series. This yields time series for the
unconstrained flows on each link, which are binned in a histogram, see Fig.\
\ref{fig:quant} for an example. It has been observed that these unconstrained
distributions generally peak around zero and have convex tails, such that a
fraction of the line capacity that would be necessary to enable the maximal
unconstrained flow is sufficient to let the flow pass through unimpededly most
of the time. The Quantile capacities are obtained by taking a certain quantile
of the unconstrained flow in each direction and setting the larger of the two as
line capacity, for each of the links. In terms of balancing energy reduction,
these have been shown to perform much better than, for instance, global scaling
of current line capacities \cite{rolando}. The resulting capacity layout is
shown in Fig.\ \ref{fig:lcall}. The costs of this layout are calculated by
taking the realistic cost estimates from Refs.\ \cite{reeds, bethany} (see Tab.\
\ref{tab:cost}) and applying them to the capacity that needs to be added on top
of what is installed today to reach the Quantile layout. The quantile for all
links is chosen to be 98.36\%, such that in total, the additional investment is
twice the cost of today's layout.

\subsection{Simulated annealing}
\label{sec:5d}

While the quantile line capacities lead to layouts that perform well in terms of
balancing reduction, they are not optimized. The problem of optimal line
capacity distribution does not take the simple form of a convex optimization.
Instead, the balancing energy as a function of line capacities appears to be
rather complicated in numerical tests, especially when the constraint of a fixed
total investment in new lines is taken into account. Simulated annealing is a
technique well suited and widely used in physics and related fields for finding
minima of such a function \cite{annealing83,annealing85}. It mimics a physical
system settling into its ground state under cooling, where it assumes a
(possibly only locally) minimal energy value. In our application, the state of
the system corresponds to a given distribution of new line capacity, a line
capacity layout or layout, for short. The energy function to be minimized is the
total balancing energy $B$. 

The system is started in a random layout. A candidate neighbor layout is
chosen by tentatively shifting approximately 100\,MW of line capacity from one
random link to another. Then, the balancing energy of the neighboring layout is
calculated, and a random decision whether to move to the candidate layout is
taken. The probability $P$ of switching from the old to the new layout is chosen
classically as:
\begin{align}
  P(B_\text{old},B_\text{new},T) = 
  \begin{cases}
    1 & \text{if } B_\text{old} > B_\text{new}\\
    \e^{-(B_\text{new}-B_\text{old})/T} & \text{otherwise}\, ,
  \end{cases}
\end{align}
where $B_\text{old}$ and $B_\text{new}$ are the balancing energy of the current
and candidate layout, respectively, and $T$ is the temperature parameter,
controlling how the space of potential layouts is scanned. If $T$ is large, the
transition probability is close to one for any candidate layout, even if
$B_\text{new}$ is much larger than $B_\text{old}$, and the system moves like a
random walk from layout to layout. For low temperatures, the acceptance
probability for shifts to layouts of higher energy goes to zero, and the system
performs an almost monotonous descend toward lower balancing energy layouts.

\begin{figure}[!th]
  \centering
  \includegraphics[width=0.49\textwidth]{\figdir/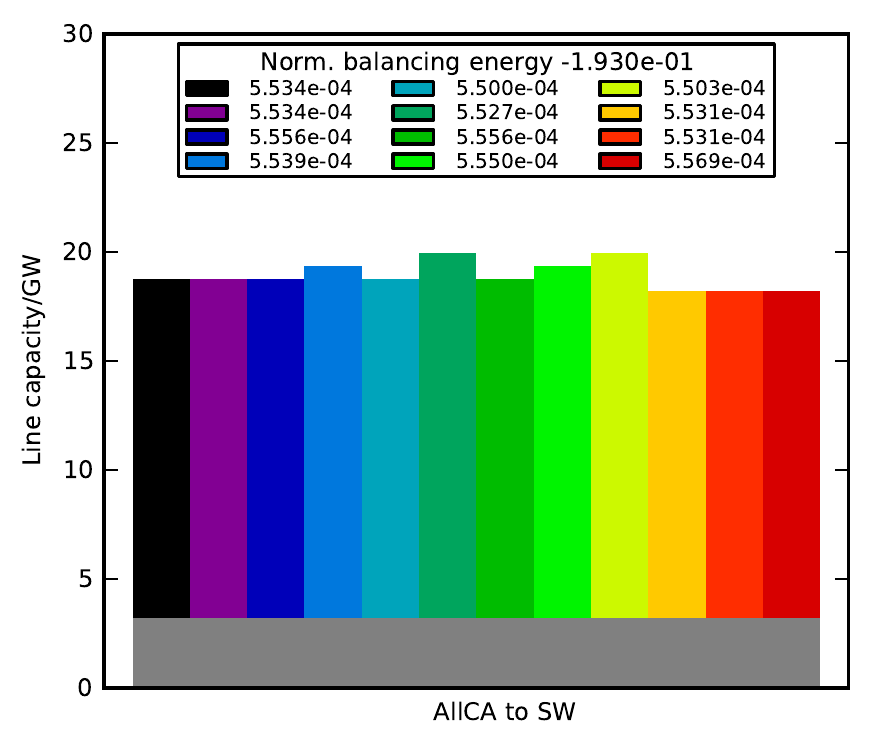}
  \caption{(Color online.) Optimized link capacities for the representative link
    between AllCA and SW, for twelve annealing runs from different start points.
    For the other links, line capacities are similarly close to each other.
    This plot shows the situation for the realistic line costs, see Sec.\
    \ref{sec:5e} for details on the costs. For the even line cost case, the
    corresponding results resemble this plot. In gray, today's line capacity is
    overlayed, which serves as a lower bound. The legend shows the resulting
    balancing energy (minus a constant offset to compare the very similar
    numbers) as a fraction of the total load, which the different line capacity
    layouts produce during the 32 years of data.
  }
  \label{fig:lcreal}
\end{figure}

In the runs presented here, $T$ is first kept at a high value to explore the
state space. From this first round, twelve start layouts per run are chosen that
lead to low balancing energies and lie sufficiently far apart. Next, annealing
is performed from these start points, linearly decreasing the temperature to
zero. To achieve better results, the best layouts from these runs are reheated
to a medium temperature and then recooled. In this way, twelve layouts with very
low balancing energy are found. Line capacities for the example link between
AllCA and SW are shown in Fig.\ \ref{fig:lcreal}.  It is visible that the
capacities almost coincide. The same holds true for the other links. Thus it
appears that there is one unique line capacity layout minimizing balancing
energy. This finding is further corroborated by looking at the spread in
balancing energy among the twelve resulting layouts, shown in Tab.\
\ref{tab:baleng}. Minima and maxima of balancing energy almost coincide,
indicating that there is a single optimal value.

The line capacities present today were enforced as lower bounds. The high
temperature was chosen such that typical transition probabilities to a higher
balancing energy layout were about 90\%, while they reached about 50\% at
reheating temperatures. The total investment was kept constant by shifting not a
fixed amount of line capacity, but line capacity of a fixed cost from one link
to the other. Due to computational limitations, the optimization was constrained
to the first two years of data.

\subsection{Line cost estimates}
\label{sec:5e}

The realistic line costs estimates are composed of different contributions:
\begin{align}
  C_l = a_l\cdot b_l \cdot C_l^\text{line} + 
  C_l^\text{substation} + 
  C_l^\text{async.}\,,
\end{align}
where $C_l^\text{line}$ are the costs of building just the line in \$$_\text{2006}$/(MW
$\cdot$ mi), $a_l$ is the line length, $b_l$ a region-specific cost multiplier
comprising differences in overall building costs, $C_l^\text{substation}$ is the
cost of substations per MW, and $C_l^\text{async.}$ is the cost of building
interties when linking asynchronous regions (the Eastern FERC regions, Western
FERC regions, and ERCOT are not synchronized with each other). Cost data come
from \cite{reeds}, and are converted to single lines, adjusted to 2006 values
and annualized as in \cite{bethany}, assuming a yearly interest rate of $i=7\%$
and a lifetime of $A=60\,\text{yrs}$:
\[
  C_l^\text{annualized} = C_l \cdot \frac{i(1+i)^A}{(1+i)^A-1}
\]
Costs are given in Tab.\ \ref{tab:cost}. Line lengths are approximated by the
distances between the geographical center points of the FERC regions they
connect. They are shown, together with current transmission capacities, in Fig.\
\ref{fig:fercs}. 

The transmission costs used here are higher than those from Ref.\
\cite{delucchi11} by an average factor of more than five. This is mainly due to the
fact that this model assumes links between different FERC regions to be spread
out over several lines, which are based on the prevalence of HVAC lines
(see \cite{reeds} for details), whereas the authors of Ref.\ \cite{delucchi11}
assume the entire transmission capacity to be aggregated in a few HVDC lines,
which are much less expensive for long-distance lines. For a fair comparison, it
has to be noted that the usage of a few HVDC lines for long distance
transmission entails more distribution lines from the end-points of these HVDC
lines which are not included in long-range transmission in Ref.\
\cite{delucchi11}, but which are partly incorporated in our approach since the
lines we are considering are distributed. Whether one or the other idea is
realized depends on how well the line build-up is coordinated and how
concentrated load centers are within the FERC regions linked.
\begin{table}[!bt]
  \centering
  \caption{Table of the costs incurred for electricity lines. $a_l$ is the line
    length in miles, calculated as described in \cite{bethany} as the distance
    between midpoints of the connected FERC regions (see also Fig.\
    \ref{fig:fercs}), $b_l$ is a region-specific, dimensionless line cost
    multiplier, $C_l^\text{line}$ is the line cost in \$$_\text{2006}$ per
    MW-mi, and $C_l^\text{async.}$ is the cost of building AC-DC-AC interties
    when linking asynchronous regions, in \$$_\text{2006}$ per kW. Not shown is
    the constant substation cost of
    $C_l^\text{substation}=16.3$\,\$$_\text{2006}$ per kW. From this input, the
    total costs in \$$_\text{2006}$ per MW shown in the last column were
    calculated. Data are taken from \cite{reeds}, adjusted to refer to lines
    (instead of regions) and annualized as described in \cite{bethany}, assuming
    an interest rate of 7\% and a lifetime of 60 years.
  }
  \label{tab:cost}
  \begin{tabular}{lrrrrrr}
    \hline
    Link & $a_l$ & $b_l$ & $C_l^\text{line}$ &
    $C_l^\text{async.}$ & total \\
    \hline
    AllCA-NW	& 520	& 2.28	& 1411	& 0.0	& 1.20$\cdot 10^{5}$ \\
    AllCA-SW	& 650	& 2.28	& 1411	& 0.0	& 1.50$\cdot 10^{5}$ \\
    ERCOT-SE	& 780	& 1.00	& 1411	& 216.4	& 9.50$\cdot 10^{4}$ \\
    ERCOT-SPP	& 460	& 1.00	& 1411	& 216.4	& 6.28$\cdot 10^{4}$ \\
    ERCOT-SW	& 645	& 1.00	& 1411	& 216.4	& 8.14$\cdot 10^{4}$ \\
    ISONE-NYISO	& 255	& 3.56	& 1129	& 0.0	& 7.42$\cdot 10^{4}$ \\
    MISO-NW	& 1045	& 1.00	& 1270	& 216.4	& 1.11$\cdot 10^{5}$ \\
    MISO-PJM	& 775	& 1.78	& 1129	& 0.0	& 1.12$\cdot 10^{5}$ \\
    MISO-SE	& 845	& 1.00	& 1270	& 0.0	& 7.76$\cdot 10^{4}$ \\
    MISO-SPP	& 520	& 1.00	& 1270	& 0.0	& 4.82$\cdot 10^{4}$ \\
    NW-SW	& 600	& 1.00	& 1411	& 0.0	& 6.15$\cdot 10^{4}$ \\
    NYISO-PJM	& 335	& 3.06	& 1129	& 0.0	& 8.37$\cdot 10^{4}$ \\
    PJM-SE	& 540	& 1.78	& 1270	& 0.0	& 8.84$\cdot 10^{4}$ \\
    SE-SPP	& 750	& 1.00	& 1411	& 0.0	& 7.66$\cdot 10^{4}$ \\
    SPP-SW	& 505	& 1.00	& 1411	& 216.4	& 6.73$\cdot 10^{4}$ \\
    \hline
  \end{tabular}
\end{table}

\subsection{Optimized line capacity layouts}
\label{sec:5f}

\begin{figure}[!bt]
  \centering
  \includegraphics[width=0.49\textwidth]{\figdir/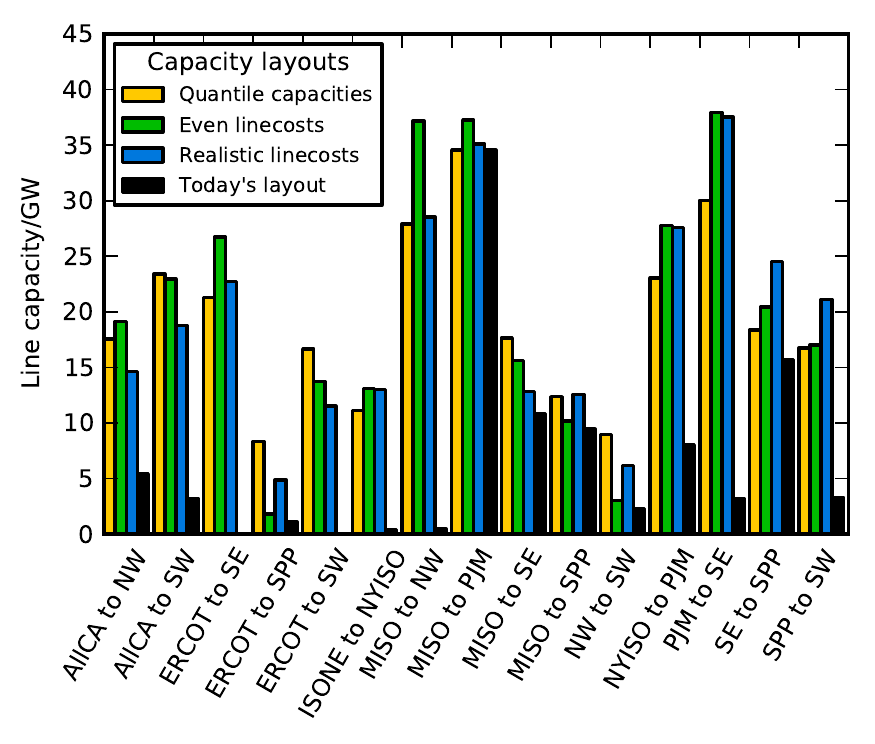}
  \caption{(Color online.) Quantile line capacity layout and two optimized
    transmission capacity layouts: if all line costs are assumed equal (Even
    line costs) and if line cost estimates as given in Tab.\ \ref{tab:cost} are
    assumed (Real line costs). In all the layouts, the total investment is set
    to twice of what is present today, and the total additional line capacity is
    chosen accordingly. For comparison, today's line capacity layout is shown as
    well. It serves as a lower bound to the grid extensions.
  }
  \label{fig:lcall}
\end{figure}
\begin{table}[!tb]
  \centering
  \caption{Percentage of total electricity consumption covered by balancing
    energy, for the three different line capacity layouts: Quantile capacities,
    cost optimal line capacities if all lines cost the same (Even layout), and
    for line cost estimates as given in Tab.\ \ref{tab:cost} (Real layout). For
    the latter two layouts, the maximal and minimal values across the twelve
    candidate layouts from the different annealing runs is also shown.
  }
  \label{tab:baleng}
  \begin{tabular}{rrll}
    \hline
    \multicolumn{2}{r}{Timespan} & 32 years & 2 years \\
    Layout & & & \\
    \hline
    Quantile &      & 19.428\% & 19.117\% \\
    \cline{2-4}
             & opt. & 19.294\% & 18.983\% \\
    Even     & min. & 19.294\% & 18.983\% \\
             & max. & 19.296\% & 18.983\% \\
    \cline{2-4}
             & opt. & 19.355\% & 19.032\% \\
    Real     & min. & 19.355\% & 19.032\% \\
             & max. & 19.356\% & 19.032\% \\
    \hline
  \end{tabular}
  \caption{Cost of different line capacity layouts when they are scaled such
    that they yield the same balancing energy reduction as the quantile capacity
    layout. Cost calculations have all been done based on the line cost
    estimates from Tab.\ \ref{tab:cost}. The first row shows the total annual
    cost. The second row contains the total annual cost normalized by the yearly
    balancing energy reduction that is achieved (compared to today's layout) by
    the new line investment. The last row shows the percentage difference in
    costs with respect to the Quantile layout.
  }
  \label{tab:linecosts}
  \begin{tabular}{rlll}
    \hline
    Layout & Quantile & Even & Real \\
    \hline
    Cost in $10^9$\$/yr  & 17.99  & 16.25  & 16.10  \\
    Cost in $\frac{\text{\$/yr}}{\text{MWh/yr}}$  & 141.02  & 127.38  & 126.22
    \\
    $\Delta$ Cost  & 0.0\%  & -9.7\%  & -10.5\%  \\
    \hline
  \end{tabular}
\end{table}
Cost-optimal line capacity layouts are calculated with simulated annealing for
two sets of prices: First, all lines are assigned the same price (an average
of the actual cost estimates from Tab.\ \ref{tab:cost}) to obtain the
Even layout. It serves as a test of the Quantile line capacities, which should
produce very similar results if performing well, as well as a sensitivity check
for the second calculation, in which we insert the line cost estimates from
Tab.\ \ref{tab:cost} to produce the Real layout. 

The Even and Real line capacity layouts resulting from simulated annealing, as
well as the Quantile line capacity layout, are shown in Fig.\ \ref{fig:lcall}.
Quantile line capacities and Even line capacities are generally similar, but
differ visibly (compare the yellow and the green bars in Fig.\ \ref{fig:lcall}).
Although they generally are within 2-5~GW of each other, deviations up to
10~GW occur. The performance of the different layouts in terms of balancing
energy reduction is shown in Tab.\ \ref{tab:baleng}. It is seen that the
balancing energy minimization with Even line costs yields lower balancing
energies than the Quantile line capacities by more than 0.1\% of the total
yearly load, or roughly 170~TWh (using 2007 load values). This means that
simulated annealing outperforms the Quantile method at the task of optimally
distributing a certain amount of additional MW in transmission capacity while
neglecting regional differences. 
\begin{figure}[!tb]
  \centering
  \begin{lpic}{\figdir/US_today_layout(3.32in)}
    \lbl{9,60;{\large \textsf{\textbf{(a)}}}}
  \end{lpic}
  \begin{lpic}{\figdir/US_real_cost_layout_diff(3.32in)}
    \lbl{9,60;{\large \textsf{\textbf{(b)}}}}
  \end{lpic}
  \begin{lpic}{\figdir/US_real_cost_layout(3.32in)}
    \lbl{9,60;{\large \textsf{\textbf{(c)}}}}
  \end{lpic}
  \includegraphics[width=0.47\textwidth]{\figdir/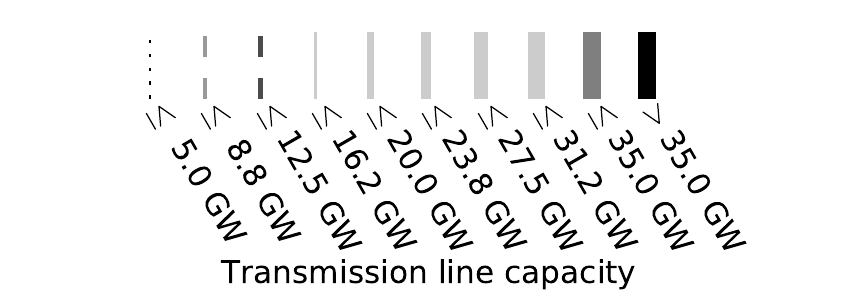}
  \caption{(Color online.) Today's layout (a) plus additional capacity to
    realize the Real cost capacity layout (b) gives the Real layout (c). Line
    thickness and style indicate transmission capacity as described in the
    legend. Node sizes and positions are not to scale.
  }
  \label{fig:ntx}
\end{figure}

When different line costs enter the game, the line capacity is redistributed to
the cheaper links, and thus balancing energy drops not as low as for the Even
cost line capacities. It should be noted, however, that the annealing method
still reduces balancing energy usage further than the Quantile capacities.

To make a cost comparison between the different capacity layouts without using
balancing energy costs which are highly complex and diverse, we scale the Real
as well as the Even layout down linearly until they lead to the same amount of
total balancing energy as the Quantile layout. The costs of the resulting
layouts are then all calculated using the line cost estimates from Tab.\
\ref{tab:cost}, and compared in Tab.\ \ref{tab:linecosts}. They are reduced by
about 10\% in both of the optimized layouts, as compared to the Quantile
capacity guess.

The additional transmission line capacity for the three cases considered here
(Quantile, Even, and Real layout) show large additions along the East Coast
(ISONE, NYISO, PJM, and SE FERC regions), West Coast (CA, NW, and SW), and
across the boundaries of the three interconnects (ERCOT to adjacent FERC
regions; MISO-NW; and SW-SPP), cf.\ Fig.\ \ref{fig:ntx}. The grid enhancement in
the NREL Futures study (for 2050), Ref.\ \cite{NREL_re_futures}, by contrast, is
mainly east-west oriented and concentrated in the middle and southwestern areas
of the US, with key additions to/from ERCOT, SE, SW, SPP, and MISO FERC regions.
These results reflect their greater emphasis on transmitting wind and solar
energy from the middle and southwestern areas of the US to large-load, adjacent
regions, while in our simulation all FERC regions are assumed to be on average
self-supplying, thus reducing the need for transmission.

Tab.\ \ref{tab:baleng} also shows the effect of calculating balancing energy
during all available years vs only relying on the first two years. While the
shift this introduces is larger than the spread between the layouts for either 2
or 32 years, it does not affect their relative distances in balancing energy
much, and in particular has no impact on their ranking.

\section{Conclusions}
\label{sec:6}

We introduced a novel, high resolution, long-term dataset for wind and solar PV
production in the contiguous US. Possible applications have been demonstrated by
calculating the optimal mix of wind and solar PV power with respect to
minimizing storage, minimizing balancing energy, and minimizing LCOE.  We showed
that by picking the right mix, the needs for storage sizes or back-up energy can
be significantly reduced. Storage energy capacities could be reduced from 30\%
to 50\% of the yearly load for a wind-only mix down to 15\% to 20\% in a
hypothetical storage-only scenario. Balancing energy could be brought down from
more than 50\% of the yearly load with solar PV only to 20\% to 25\%, depending
on the transmission grid. Furthermore, we investigated the influence of
installation and operation costs on the optimal wind-solar mix and showed how
sensitive this mix is to relative prices, highlighting the need for reliable
price predictions. Again, picking the optimal mix reduced expenses (here LCOE),
by a factor of almost two. Interestingly, when taking effects of surplus
production that is lost to the system into account, LCOE differences between
wind and solar are seen to play an increasingly minor role with rising renewable
penetration. The LCOE-optimized mix is instead driven by the avoidance of losses
from surplus production, which becomes more important than the installation
costs. This can be read as an indication that it pays in the long term to
maintain a varied technology mix in spite of different installation costs.

As a second application, we calculated optimized transmission grid extensions,
showing the importance of a careful numerical optimization of the grid's
capacity layout. Introducing transmission line costs helped to integrate
regional differences such as different labor and equipment costs and different
line lengths. We showed, however, that the improvement over an ad-hoc assumption
of uniform line costs per MW was not very significant, resulting in 0.8\% higher
costs for the same reduction in balancing energy. This also highlights that our
analysis is relatively insensitive to transmission line costs. The main
improvement of the study was introducing simulated annealing techniques in
optimizing the transmission line capacities, which led to a cost reduction by
more than 10\% compared to simpler approaches.

\section*{Acknowledgments}

SB gratefully acknowledges financial support from O.\ and H.\ St\"ocker as well as
M.\ and H.\ Puschmann, BAF from a National Defense Science and Engineering Graduate
(NDSEG) fellowship and a National Science Foundation (NSF) graduate fellowship,
and GBA from DONG Energy and the Danish Advanced Technology Foundation.
Furthermore, we thank Anders A.\ S\o ndergaard for helpful and constructive
discussions.

\bibliographystyle{unsrt}
\bibliography{literatur}

\begin{appendix}

\section{Wind conversion}
\label{app:wind}

\subsection{Original approach}

In a first attempt, the wind conversion was done with the Aarhus RE atlas as
described in \cite{anders}. Summarized briefly, the wind speed at hub height was
extrapolated from measurements of wind speeds at 10\,m height $u$(10\,m) via
\begin{align}
  u(H) = u(10{\rm m})\cdot\frac{\ln(H/z_0)}{\ln(10{\rm m}/z_0)},
\end{align}
where $u$ is the wind speed as a function of height, $H$ is the hub height, and
$z_0$ is the surface roughness. The wind speed was in turn converted into
turbine power output with the help of power curves as can be found in data sheets
for wind turbines. Specifically, we used data for the model Vestas V90~3\,MW with
80\,m hub height onshore and for Vestas V164~7\,MW with 100\,m hub height offshore.

When evaluating the data, it became apparent that this approach significantly
underestimates the wind power potential, see Fig.\ \ref{fig:cmp}a and c.

\subsection{Modified approach}

In order to fix this issue, the methods described in Refs.\ \cite{BJ2010,BJ2011}
were applied. The main reason that wind power potential is underestimated
onshore is that the spatial fluctuations in wind speed due to surface roughness
and orography (hills and valleys) is neglected when working with spatially
averaged wind speeds. However, these fluctuations contribute to the mean wind
energy density, as the following calculation shows:
\begin{align}
  \frac{2\langle e\rangle}{\rho} = \langle u^3\rangle 
  = \langle(\overline{u}+u')^3\rangle
  \approx \overline{u}^3 + 3\overline{u}\langle u'^2\rangle
  = \overline{u}^3 + 3\overline{u}\sigma^2
  \label{eq:ed}
\end{align}
$e$ is the wind energy density, $\rho$ is the mass per volume air density
(assumed constant), $\overline{u}$ is the mean wind speed, $u'$ are the spatial
fluctuations around $\overline{u}$ such that $u=\overline{u}+u'$, and $\sigma$
is the standard deviation of $u'$.

Interestingly, the wind speed fluctuations due to inhomogeneous terrain can be
characterized in terms of its surface roughness and orography, yielding
\begin{align}
  \sigma = \sqrt{\sigma_\text{oro}^2+\sigma_\text{rough}^2}.
\end{align}
The details of this connection are given below. The resulting $\sigma$ is then
converted to an effective increase in wind speed, which leads to an increased
power output.

With the additional assumption that the distribution of $u'$ is Gaussian, even
the effect of siting can be modeled in a simplified way: Wind turbines would
primarily be placed in good spots, that is, where $u'$ is larger than some
threshold value $p$. This leads to mean energy densities of
\begin{align}
  \frac{2\langle e\rangle}{\rho} = \langle u^3\rangle 
  = \langle(\overline{u}+u'\cdot\theta(u' > p))^3\rangle\, ,
  \label{eq:p}
\end{align}
where $\theta$ is a cutoff function (1 if the condition is true, 0 otherwise).

The methods to calculate $\sigma_\text{oro}$ and $\sigma_\text{rough}$ yield
standard deviations normalized by mean wind speed. Therefore, they first have to
be "de-normalized" by multiplication with the corresponding grid cell's mean
wind speed (the same that was used in the normalization). Since the mean wind
speeds used for roughness and orography differ slightly, we use these two
different speeds before adding them together, see Eq.\ \eqref{eq:denorm}. Next,
the effective increase in wind speed at hub height $H$ has to be calculated:
\begin{align}
  \frac{2 \Delta e(H)}{\rho} &= (\overline{u}(H)+\Delta u)^3-\overline{u}(H)^3
  \stackrel{!}{=} 3\overline{u}(H)\sigma^2\\
  \Rightarrow \Delta u &= \overline{u}(H)\left(\sqrt[3]{1+3 \sigma_\text{norm}^2} -
  1\right) \\
  \text{with }
  \sigma_\text{norm}^2 &= \frac{\sigma_\text{oro,norm}^2\cdot
  \overline{u}_\text{oro}^2
  + \sigma_\text{rough,norm}^2\cdot
  \overline{u}_\text{rough}^2}{\overline{u}^2}
  \label{eq:denorm} 
\end{align}
If only the best spots in each grid cell are considered as potential wind sites
(i.e.\ we work with Eq. \eqref{eq:p} instead of \eqref{eq:ed}), the
corresponding wind speed correction takes the following form:
\begin{align}
  \Delta u =
  \frac{\overline{u}}{N}\Bigg\{&\cdot 3\cdot\frac{1}{\sqrt{2\pi}} 
  \sigma_\text{norm}\e^{-p^2/(2\sigma_\text{norm}^2)} \nonumber\\
  &+\, 3\cdot\frac{1}{2}\sigma_\text{norm}^2
  \left[1-\frac{2p}{\sqrt{2\pi\sigma_\text{norm}^2}} \cdot \right.
    \label{eq:wildsigmas} \\
    & \left. \qquad \qquad \qquad \e^{-p^2/(2\sigma_\text{norm}^2)}
      \erf\left(\frac{p}{\sqrt{2\sigma_\text{norm}^2}}\right)
  \right] \nonumber\\
  &+ \sqrt{\frac{2}{\pi}}\sigma_\text{norm}^3
  (1+\frac{p^2}{2\sigma_\text{norm}^2})\e^{-p^2/(2\sigma_\text{norm}^2)}\Bigg\}\, ,
  \nonumber \\
  \text{with }
  N =& \frac{1}{2} \left( 1-\erf\left(\frac{p}{\sqrt{2\sigma_\text{norm}^2}}
  \right) \right) \, ,
\end{align}
where $\erf$ is the error function. Wind is upscaled first, before the
correction is added, so the final formula reads
\begin{align}
  u_\text{corrected}(H) = u(H)+\Delta u\,.
\end{align}

\subsubsection{Surface roughness}

We closely follow \cite{BJ2010}. As roughness dataset, we use the land cover
atlas from the national land cover database for the US from \cite{nlcd2006}. It
gives land use classes with a resolution of 30\,m. The land use classes are
converted to surface roughness lengths using Tab.\ \ref{tab:sr} in
\ref{app:roughness}. From these data, a contribution to wind speed fluctuations
for each grid cell is calculated. Then, the surface layer friction velocity,
$u_\star$, is obtained by numerically inverting the geostrophic drag law
\begin{align}
  G=\frac{u_\star}{\kappa}\sqrt{\left(\ln\frac{u_\star}{f
  z_0}-A\right)^2+B^2}\,.
\end{align}
The geostrophic wind $G$ is assumed to be 10\,m/s, $\kappa\approx0.4$ is the
Karman constant, $f\approx10^{-4}$ is the Coriolis frequency (for latitudes of
the contiguous US), and $A$ and $B$ are dimensionless parameters which take the
values $1.8$ and $4.5$ respectively, for stable atmospheric conditions (see
e.g.\ \cite{petersen1997}). $z_0$ is again the surface roughness length.

From $u_\star$, the wind speed at height $H$ is calculated by using again a
logarithmic scaling law:
\begin{align}
  u(H) = \frac{u_\star}{\kappa}\ln\frac{H}{z_0}
\end{align}

\begin{figure}[!htb]
  \centering
  \includegraphics[width=0.49\textwidth]{\figdir/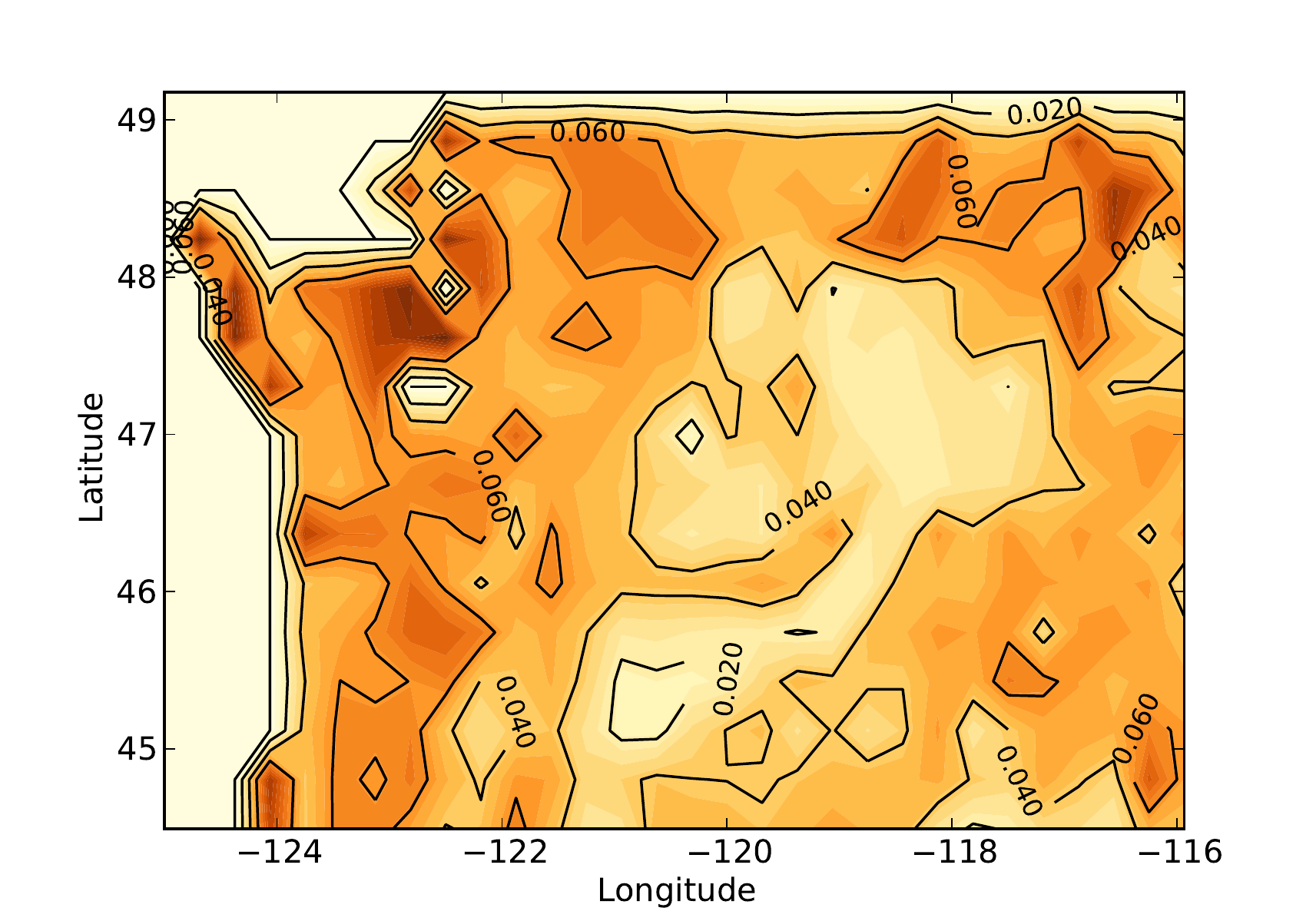}
  \caption{(Color online.) Corrections from surface roughness effects from our
    conversion for the Columbia Gorge region, to compare to Fig.\ 6 from Ref.\
    \cite{BJ2011}. Note that the cutout in this image is not perfectly aligned
    with the reference image.
  }
  \label{fig:roughness}
\end{figure}

According to \cite{BJ2010}, the contribution to the wind speed standard
deviation from this can be calculated as the standard deviation of $u(H)$ (for
all points within one grid cell). Normalized by the mean wind speed of the grid
cell, Fig.~6 from Ref.\ \cite{BJ2011} shows their corrections for the Columbia
Gorge region. Our corresponding data are shown in Fig.\ \ref{fig:roughness},
scaled by a factor of $0.6$. The agreement seems reasonable. We believe that the
need for scaling our results down to match theirs arises from the higher spatial
resolution of our data; they use surface roughness input data with a resolution
of 500\,m.

\subsubsection{Orography}

\begin{figure}[!bt]
  \centering
  \includegraphics[width=0.49\textwidth]{\figdir/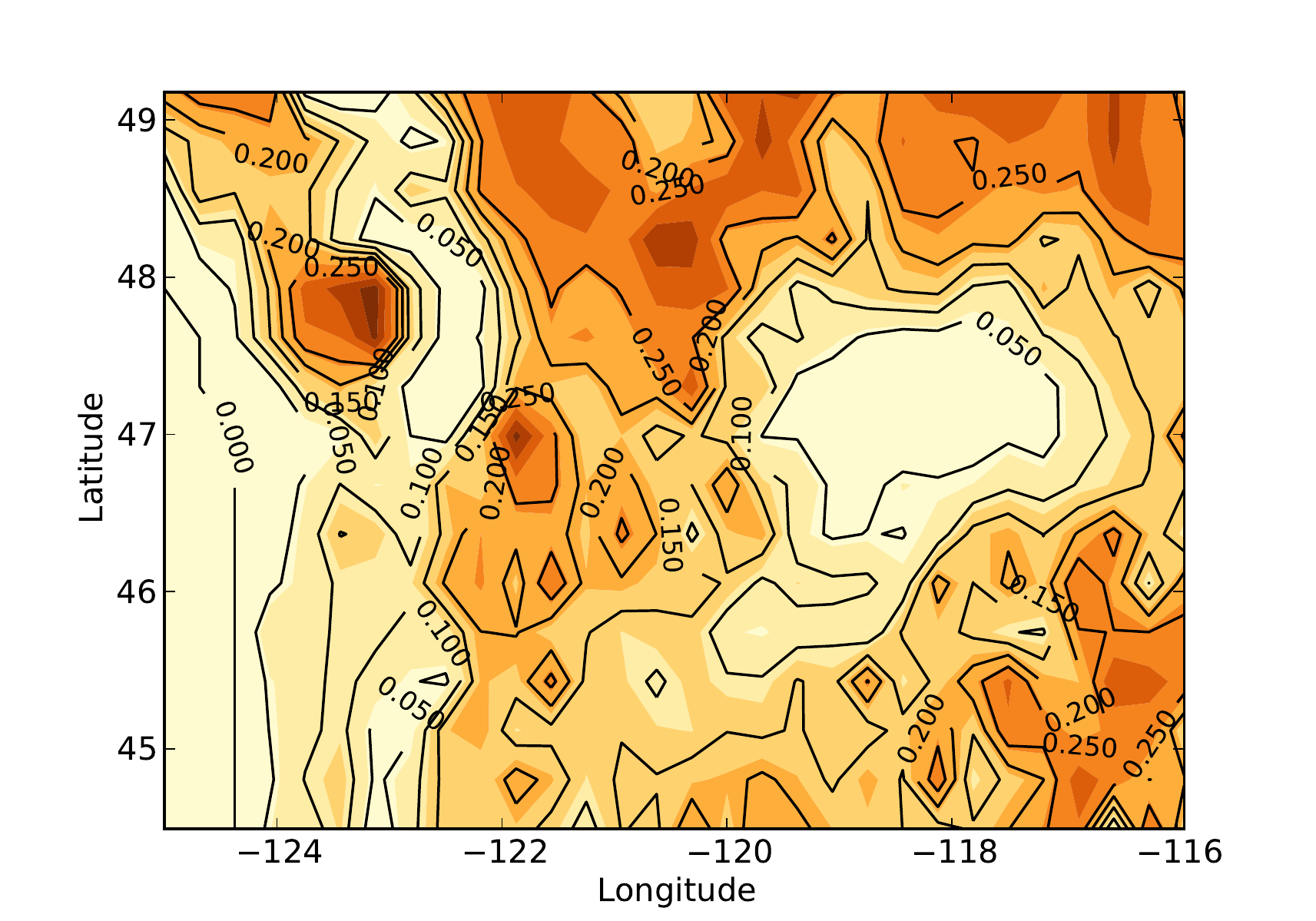}
  \caption{(Color online.) Corrections from orography effects from our
    conversion for the Columbia Gorge region, to be compared to Fig.\ 5 from
    Ref.\ \cite{BJ2011}. Note that the cutouts in the two figures are not
    perfectly aligned.
  }
  \label{fig:orography}
\end{figure}

For orography, the ideas of \cite{BJ2010} are only roughly applied, since the
full details of their implementation are, to our knowledge, unpublished. The
basic idea is that wind speed-up due to the topography of a landscape should be
proportional to its unevenness. As a measure of that, the standard deviation of
the elevation, as reported in \cite{esri2005}, is employed. The data has a
resolution of 1000\,m.  When scaled by $\frac{1}{3}\cdot\frac{1}{8\pi^2}$ (first
factor from by-eye fit, second from \cite{BJ2010}) and normalized by the mean
wind speed, the agreement with \cite{BJ2011} is reasonable, compare Fig.\
\ref{fig:orography} to Fig.\ 5 from Ref.\ \cite{BJ2011}.

\subsection{Effect of the corrections}

\begin{figure*}[p]
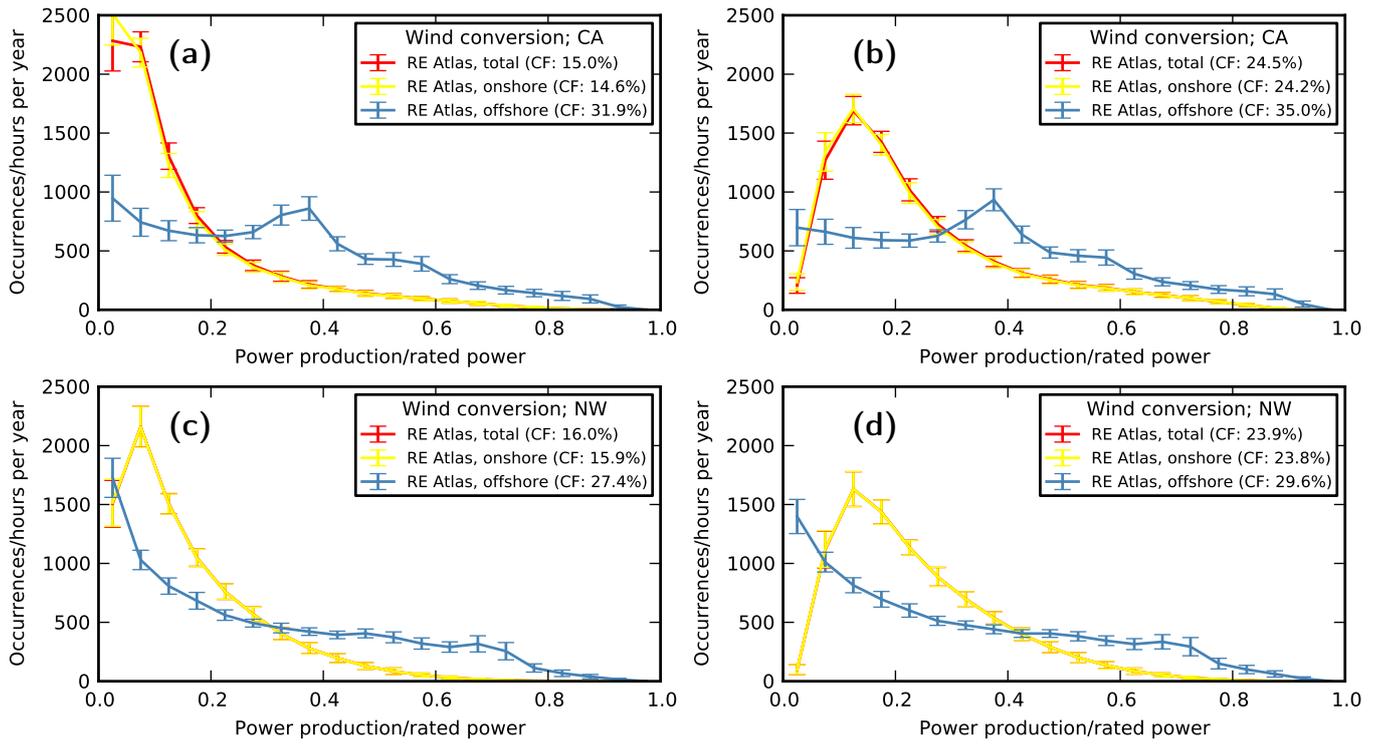

  \centering
  \begin{lpic}{\figdir/wind_solar_statistics_CA_conversion_old(3.5in)}
    \lbl{24,41;{\large \textsf{\textbf{(a)}}}}
  \end{lpic}
  \begin{lpic}{\figdir/wind_solar_statistics_CA_conversion(3.5in)}
    \lbl{24,41;{\large \textsf{\textbf{(b)}}}}
  \end{lpic}
  \begin{lpic}{\figdir/wind_solar_statistics_NW_conversion_old(3.5in)}
    \lbl{24,41;{\large \textsf{\textbf{(c)}}}}
  \end{lpic}
  \begin{lpic}{\figdir/wind_solar_statistics_NW_conversion(3.5in)}
    \lbl{24,41;{\large \textsf{\textbf{(d)}}}}
  \end{lpic}
  \caption{(Color online.) The conversions with the RE atlas with and without
    the corrections are compared. Wind power output histograms without the
    correction ((a) and (c)), and with the correction ((b) and (d)), including
    the best $20\%$ of the area within each grid cell, for the California region
    ((a) and (b)), and the NW FERC region ((c) and (d)). It is obvious that
    especially onshore, wind power output is systematically underestimated
    without the correction.
  }
  \label{fig:cmp}
\end{figure*}

\begin{figure*}[p]
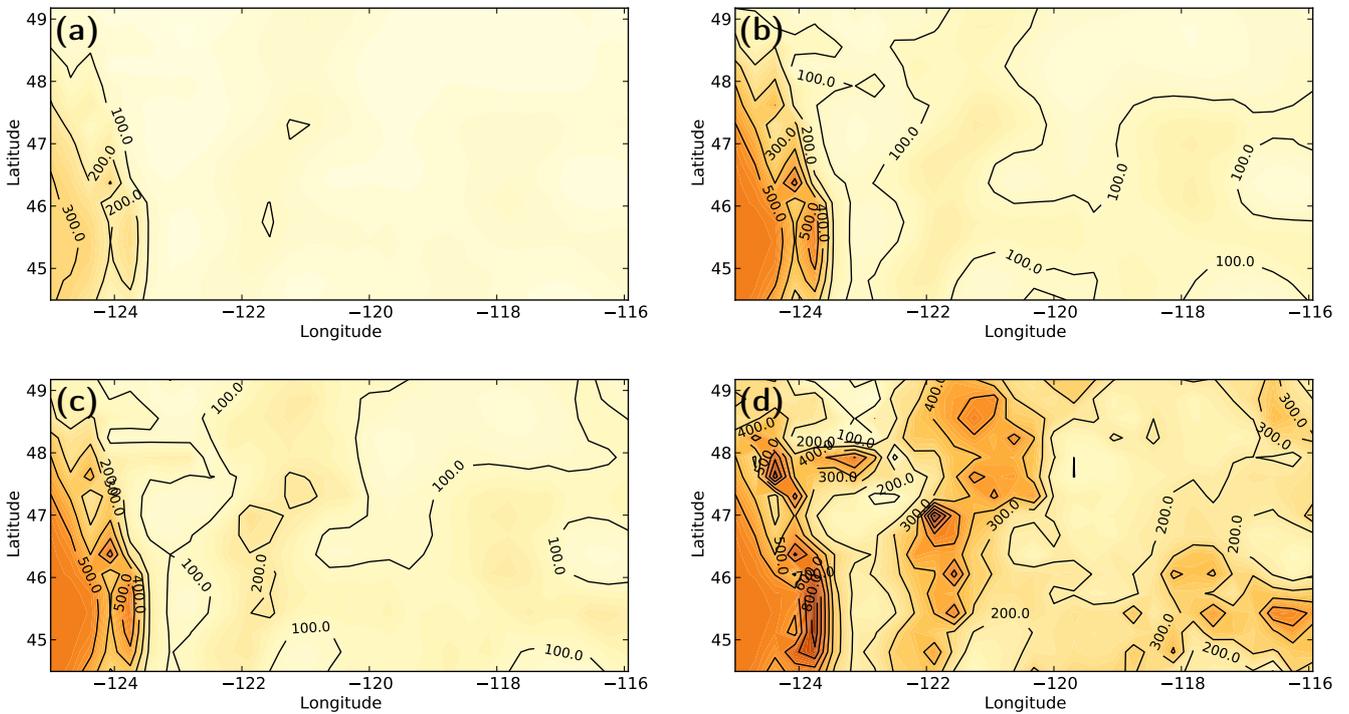

  \centering
  \begin{lpic}{\figdir/mean_wind(3.5in)}
    \lbl{23,85;{\large \textsf{\textbf{(a)}}}}
  \end{lpic}
  \begin{lpic}{\figdir/mean_cubed_wind(3.5in)}
    \lbl{23,85;{\large \textsf{\textbf{(b)}}}}
  \end{lpic}
  \begin{lpic}{\figdir/mean_cubed_wind_plus_sigma(3.5in)}
    \lbl{23,85;{\large \textsf{\textbf{(c)}}}}
  \end{lpic}
  \begin{lpic}{\figdir/mean_cubed_wind_plus_sigma_N_sigma_1.28(3.5in)}
    \lbl{23,85;{\large \textsf{\textbf{(d)}}}}
  \end{lpic}
  \caption{(Color online.) Plots analogous to Figs.\ 7-10 in \cite{BJ2011} for
    the Columbia Gorge region. Mean wind energy density in 50\,m height based on
    (a) mean wind speed cubed, (b) mean of the cube of the wind speeds, (c) as
    (b) plus the corrections, including all sites within a grid cell, (d) as
    (c), but including only the windiest 10\% of the wind speed distribution.
    Color coding is the same in all plots.
  }
  \label{fig:bj710}
\end{figure*}

The results of the procedure described above with a choice of $p=0.84\sigma$
(corresponding to picking the best $20\%$ of the area in each grid cell as
eligible for wind farms) is shown in Fig.\ \ref{fig:cmp}b and d. Another way to
illustrate the effects of the wind speed correction is by looking at plots
analogous to Fig.\ 7-10 from \cite{BJ2011}, which show the expected energy yield
of each site as calculated from mean wind speed, mean cubed wind speed, mean
cubed wind speed taking the surface roughness and orography corrections into
account, and mean wind speed taking the corrections as well as siting effects
into account. For our data, these are given in Fig.\ \ref{fig:bj710}a-d. The
qualitative agreement is reasonable.

\onecolumn

\section{Land use to roughness table}
\label{app:roughness}

{\small
  \centering
  \begin{longtable}{rp{0.75\textwidth}r}
  \caption{Land use - surface roughness relations used for the conversion from
    the NLCD atlas to surface roughness. Some classes not shown because they
    only apply to Alaska.
  } \\
  \label{tab:sr}
  USGS ID & Class & Roughness/m \\
  \endfirsthead
  USGS ID & Class & Roughness/m \\
  \endhead
  11 & Open Water - All areas of open water, generally with less than 25\% cover
  or vegetation or soil & 0.0002 \\
  12 & Perennial Ice/Snow - All areas characterized by a perennial cover of ice
  and/or snow,generally greater than 25\% of total cover. & 0.005 \\
  21 & Developed, Open Space - Includes areas with a mixture of some constructed
  materials, but mostly vegetation in the form of lawn grasses. Impervious
  surfaces account for less than 20 percent of total cover. These areas most
  commonly include large-lot single-family housing units, parks, golf courses,
  and vegetation planted in developed settings for recreation, erosion control,
  or aesthetic purposes. & 0.5 \\
  22 & Developed, Low Intensity -Includes areas with a mixture of constructed
  materials and vegetation. Impervious surfaces account for 20-49 percent of
  total cover. These areas most commonly include single-family housing units.
  & 1 \\
  23 & Developed, Medium Intensity - Includes areas with a mixture of
  constructed materials and vegetation. Impervious surfaces account for 50-79
  percent of the total cover. These areas most commonly include single-family
  housing units. & 1 \\
  24 & Developed, High Intensity - Includes highly developed areas where people
  reside or work in high numbers. Examples include apartment complexes, row
  houses and commercial/industrial. Impervious surfaces account for 80 to 100
  percent of the total cover. & 2 \\
  31 & Barren Land (Rock/Sand/Clay) - Barren areas of bedrock, desert pavement,
  scarps, talus, slides, volcanic material, glacial debris, sand dunes, strip
  mines, gravel pits and other accumulations of earthen material. Generally,
  vegetation accounts for less than 15\% of total cover. & 0.03 \\
  41 & Deciduous Forest  - Areas dominated by trees generally greater than 5
  meters tall, and greater than 20\% of total vegetation cover. More than 75
  percent of the tree species shed foliage simultaneously in response to
  seasonal change. & 1 \\
  42 & Evergreen Forest - Areas dominated by trees generally greater than 5
  meters tall, and greater than 20\% of total vegetation cover. More than 75
  percent of the tree species maintain their leaves all year. Canopy is never
  without green foliage. Enumerated\_Domain\_Value\_Definition\_Source: NLCD
  Legend Land Cover Class Descriptions & 1 \\
  43 & Mixed Forest - Areas dominated by trees generally greater than 5 meters
  tall, and greater than 20\% of total vegetation cover. Neither deciduous nor
  evergreen species are greater than 75 percent of total tree cover. & 0.5 \\
  52 & Shrub/Scrub - Areas dominated by shrubs; less than 5 meters tall with
  shrub canopy typically greater than 20\% of total vegetation. This class
  includes true shrubs, young trees in an early successional stage or trees
  stunted from environmental conditions. & 0.25 \\
  71 & Grassland/Herbaceous - Areas dominated by grammanoid or herbaceous
  vegetation, generally greater than 80\% of total vegetation. These areas are
  not subject to intensive management such as tilling, but can be utilized for
  grazing. & 0.1 \\
  81 & Pasture/Hay  - Areas of grasses, legumes, or grass-legume mixtures
  planted for livestock grazing or the production of seed or hay crops,
  typically on a perennial cycle. Pasture/hay vegetation accounts for greater
  than 20 percent of total vegetation. & 0.03 \\
  82 & Cultivated Crops - Areas used for the production of annual crops, such as
  corn, soybeans, vegetables, tobacco, and cotton, and also perennial woody
  crops such as orchards and vineyards. Crop vegetation accounts for greater
  than 20 percent of total vegetation. This class also includes all land being
  actively tilled. & 0.1 \\
  90 & Woody Wetlands - Areas where forest or shrub land vegetation accounts for
  greater than 20 percent of vegetative cover and the soil or substrate is
  periodically saturated with or covered with water. & 0.25 \\
  95 & Emergent Herbaceous Wetlands - Areas where perennial herbaceous
  vegetation accounts for greater than 80 percent of vegetative cover and the
  soil or substrate is periodically saturated with or covered with water. & 0.03
  \end{longtable}
}
\end{appendix}

\end{document}